\documentclass[12pt,english,draftcls, one column]{IEEEtran}

\usepackage[T1]{fontenc}
\usepackage[latin9]{inputenc}
\usepackage{float}
\usepackage{amsmath}
\usepackage{amsthm}
\usepackage{amssymb}
\usepackage{graphicx}
\usepackage{stfloats}
\usepackage{cite}
\usepackage{xcolor}
\usepackage{bbm}
\usepackage{multirow}
\usepackage{multicol}
\usepackage{subfig}

\makeatletter

\floatstyle{ruled}
\newfloat{algorithm}{tbp}{loa}
\providecommand{\algorithmname}{Algorithm}
\floatname{algorithm}{\protect\algorithmname}

\theoremstyle{plain}

\theoremstyle{plain}

\IEEEoverridecommandlockouts
\usepackage{cite}
\usepackage{amsfonts}\usepackage{algorithmic}
\usepackage{textcomp}
\usepackage{etoolbox}

\def\BibTeX{{\rm B\kern-.05em{\sc i\kern-.025em b}\kern-.08em
    T\kern-.1667em\lower.7ex\hbox{E}\kern-.125emX}}

\patchcmd{\@maketitle}
  {\addvspace{0.5\baselineskip}\egroup}
  {\addvspace{-1\baselineskip}\egroup}
  {}
  {}

\makeatother

\providecommand{\propositionname}{Proposition}
\providecommand{\theoremname}{Theorem}

\begin{document}

\title{\huge Accelerating Multi-UAV Collaborative Sensing Data Collection: A Hybrid TDMA-NOMA-Cooperative Transmission in Cell-Free MIMO Networks}

\author{Eunhyuk Park, \textit{Graduate Student Member}, \textit{IEEE}, Junbeom Kim, \textit{Member}, \textit{IEEE}, \\Seok-Hwan Park, \textit{Senior Member}, \textit{IEEE}, Osvaldo Simeone, \textit{Fellow}, \textit{IEEE}, \\and Shlomo Shamai (Shitz), \textit{Life Fellow}, \textit{IEEE} \thanks{


The work of E. Park and S.-H. Park was partially supported by the National Research Foundation (NRF) of
Korea, funded by the MOE under Grant 2019R1A6A1A09031717 and MSIT
under Grant RS-2023-00238977, as well as by research funds of Jeonbuk National University in 2024. The work of J. Kim was supported by the research grant of the Gyeongsang National University in 2023. The work of O. Simeone was supported by European Union's Horizon Europe project CENTRIC (101096379), by the Open Fellowships of the EPSRC (EP/W024101/1), and by the EPSRC project (EP/X011852/1). The work of S. Shamai was supported by the German Research Foundation (DFG) via the German-Israeli Project Cooperation (DIP), under Project SH 1937/1-1.  \textit{(Corresponding authors: Junbeom Kim and Seok-Hwan Park.)}

E. Park and S.-H. Park are with the Division of Electronic Engineering, Jeonbuk
National University, Jeonju, Korea (email: uool$\_$h@jbnu.ac.kr, seokhwan@jbnu.ac.kr).

J. Kim is with the Department of AI Information Engineering, Gyeongsang National University, Jinju, Korea (email: junbeom@gnu.ac.kr).

O. Simeone is with the King's Communications, Learning $\&$ Information Processing (KCLIP) lab within the Centre for Intelligent Information Processing Systems (CIIPS), Department of Engineering, King's College London, London WC2R 2LS, U.K. (email: osvaldo.simeone@kcl.ac.uk).

Shlomo Shamai (Shitz) is with the Department of Electrical and Computer Engineering, Technion, Haifa 3200003, Israel (e-mail: sshlomo@ee.technion.ac.il).

Copyright (c) 2024 IEEE. Personal use of this material is permitted. However, permission to use this material for any other purposes must be obtained from the IEEE by sending a request to pubs-permissions@ieee.org.}
}
\maketitle
\begin{abstract}

This work investigates a collaborative sensing and data collection system in which multiple unmanned aerial vehicles (UAVs) sense an area of interest and transmit images to a cloud server (CS) for processing. To accelerate the completion of sensing missions, including data transmission, the sensing task is divided into individual private sensing tasks for each UAV and a common sensing task that is executed by all UAVs to enable cooperative transmission.
Unlike existing studies, we explore the use of an advanced cell-free multiple-input multiple-output (MIMO) network, which effectively manages inter-UAV interference. To further optimize wireless channel utilization, we propose a hybrid transmission strategy that combines time-division multiple access (TDMA), non-orthogonal multiple access (NOMA), and cooperative transmission.
The problem of jointly optimizing task splitting ratios and the hybrid TDMA-NOMA-cooperative transmission strategy is formulated with the objective of minimizing mission completion time. Extensive numerical results demonstrate the effectiveness of the proposed task allocation and hybrid transmission scheme in accelerating the completion of sensing missions.

\end{abstract}


\theoremstyle{theorem}
\newtheorem{theorem}{Theorem}
\theoremstyle{proposition}
\newtheorem{proposition}{Proposition}
\theoremstyle{lemma}
\newtheorem{lemma}{Lemma}
\theoremstyle{corollary}
\newtheorem{corollary}{Corollary}
\theoremstyle{definition}
\newtheorem{definition}{Definition}
\theoremstyle{remark}
\newtheorem{remark}{Remark}

\section{Introduction}

Unmanned Aerial Vehicle (UAV)-based sensing is promised to become a leading sensing technology in the future due to its extensive coverage area and flexible observation capabilities \cite{Wu:JSAC21,Gu:CC23,Banafaa:ACC24}.
This system offers rapid mission execution as UAVs can transmit sensory data to ground base stations (BSs) through line-of-sight (LoS) wireless links, eliminating the need to return to the BSs.
Additionally, UAVs excel in sensing areas inaccessible to terrestrial vehicles due to their increased heights and less signal blockage \cite{Wu:JSAC21,Gu:CC23,Banafaa:ACC24}, further enhancing their utility.
References \cite{Zhang:CL18} and \cite{Liu:WCL23} explored the optimization of UAV trajectory and transmit power to maximize energy efficiency.
In reference \cite{Meng:TVT19}, the authors addressed the joint design of UAV trajectory, UAV-BS association, and sensing order with the aim of minimizing the overall mission completion time.
Reference \cite{Xu:WCSP23} investigated a UAV-assisted integrated sensing and communication (ISAC) system, focusing on the joint optimization of UAV trajectory, sensing and communication time duration, and UAV power allocation.
References \cite{Zhou:IOT24} and \cite{Liu:TCCN24} proposed UAV-aided joint communication, sensing, and computing frameworks, leveraging mobile edge computing capability at BSs.
Furthermore, UAV sensing for cellular inter-cell interference coordination (ICIC) was investigated in \cite{Mei:WC21, Burhanuddin:TCCN23}. In particular, reference \cite{Mei:WC21} highlighted the spectrum sensing capabilities of UAVs across wide coverage areas, which can be leveraged to overcome the limitations of traditional terrestrial ICIC techniques.
The synergetic benefits of integrating UAV-based networks with reconfigurable intelligent surfaces (RISs), along with corresponding design strategies, were explored in \cite{Pang:WC21, Zhang:TWC22, Lima:ComSoc22}.
Specifically, the impact of operating RISs on the ground to assist air-to-ground communication between UAVs and terrestrial BSs was studied in \cite{Pang:WC21} and \cite{Zhang:TWC22}, while the potential of aerial RISs mounted on UAVs was discussed in \cite{Pang:WC21} and \cite{Lima:ComSoc22}.
Given the complexity of integrating UAV sensing techniques with other key enablers for 6G wireless systems, the application of artificial intelligence (AI) techniques is crucial for practical optimization.
The benefits of utilizing AI for UAV cognitive sensing tasks, such as object detection in aerial scenes captured by UAVs, were examined in \cite{Jain:Netw21}, while reference \cite{Lahmeri:ComSoc21} provided an extensive overview of potential AI applications for UAV-based wireless networks.

UAV-assisted communication systems are a key component of non-terrestrial networks, which also encompass satellite and high-altitude platform (HAP) systems \cite{Zin:TAES22, Lin:TC21, An:TWC, Lin:IOT21}.
In \cite{Zin:TAES22} and \cite{Lin:TC21}, beamforming design problems were explored for satellite and terrestrial integrated networks (STINs), with \cite{Zin:TAES22} focusing on the use of RISs and \cite{Lin:TC21} addressing the criterion of maximizing secrecy-energy efficiency.
Reference \cite{An:TWC} discussed beamforming design for simultaneous wireless information and power transfer (SWIPT) in RIS-assisted HAP networks, where each receiver is equipped multi-layer refracting RISs. Additionally, \cite{Lin:IOT21} investigated multicast beamforming design for satellite and aerial integrated networks (SAINs), employing rate-splitting multiple access (RSMA) for UAV transmission.

\begin{figure}
\centering\includegraphics[width=1\linewidth]{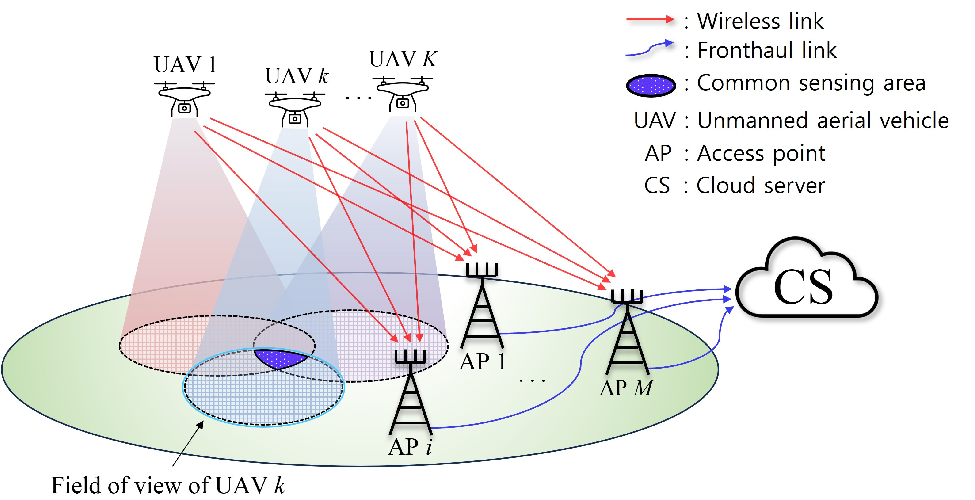}
\caption{\small Illustration of a collaborative sensing system with multiple multi-antenna UAVs reporting to a CS through distributed multi-antenna APs.} \label{fig:system-model}
\vspace{-6mm}
\end{figure}

While references \cite{Zhang:CL18, Meng:TVT19, Xu:WCSP23, Zhou:IOT24, Liu:TCCN24} focused on the UAV sensing and transmission systems with a single UAV,
there has been a notable shift towards investigating multi-UAV cooperative sensing systems. As illustrated in Fig. \ref{fig:system-model}, these systems involve multiple UAVs collaborating to conduct the sensing task \cite{Hu:IOT19, Chen:TVT20, Wu:TC20}.
However, due to the independence of sensory data signals from different UAVs, orthogonal multiple access techniques such as time-division multiple access (TDMA) were considered to avoid interference among UAVs' transmissions.
In contrast, in \cite{Meng:TWC23}, the authors proposed the partial utilization of a common sensing task executed by all UAVs. Despite the inefficiency stemming from the repeated execution of the same task among UAVs, the shared sensory data can be possessed by all UAVs, thereby reducing uplink transmission time through cooperative transmission.
Since the presence of a common sensing task yields conflicting impacts on the overall mission completion time, the work \cite{Meng:TWC23} addressed the problem of finding the optimal portions of common and private tasks.

\begin{table*}[t]
\caption{\small Comparison of existing UAV sensing and transmission literature.}
\centering
\renewcommand{\arraystretch}{1.0}
\begin{tabular}{|c||c|c||c||c|c||c|c|}
    \hline
    \multirow{2}{*}{Papers} & \multicolumn{2}{c||}{Number of nodes} & \multirow{2}{*}{Task allocation} & \multicolumn{2}{c||}{Number of antennas} & \multicolumn{2}{c|}{Transmission}  \\
    \cline{2-3} \cline{5-8} & UAV & AP & & UAV & AP & Network & Scheme \\
    \hline \hline
    \cite{Zhang:CL18} &  single &  single & - & single & single & cellular & P2P \\
    \hline
    \cite{Liu:WCL23} &  single &  single & - & single & single & cellular & P2P \\
    \hline
    \cite{Meng:TVT19} &  single &  single & - & single & single & cellular & P2P \\
    \hline
    \cite{Xu:WCSP23} &  single &  single & - & single & single & cellular & P2P \\
    \hline
    \cite{Zhou:IOT24} &  single &  single & - & single & single & cellular & P2P \\
    \hline
    \cite{Liu:TCCN24} &  single &  single & - & multiple & multiple & cellular & P2P \\
    \hline
    \cite{Hu:IOT19} &  multiple &  single & exclusive & single & multiple & cellular & orthogonal \\
    \hline
    \cite{Chen:TVT20} &  multiple &  single & exclusive & multiple & multiple & cellular & TDMA \\
    \hline
    \cite{Wu:TC20} & multiple &  single & exclusive & single & single & cellular & TDMA \\
    \hline
    \cite{Meng:TWC23} &  multiple & single & exclusive/overlap. & single & single & cellular & TDMA/coop. \\
    \hline
    This work & multiple & multiple & exclusive/overlap. & multiple & multiple & cell-free & TDMA/NOMA/coop. \\
    \hline
\end{tabular}
~

~

\end{table*}

While reference \cite{Meng:TWC23} highlighted the advantages of employing a common sensing task to reduce wireless transmission of sensory data, it primarily focused on TDMA-based transmission of private sensory data within a simplistic uplink network served by a single BS, with both the UAVs and the BS equipped with a single antenna.
Building on this background, in this work, we explore the impact of incorporating a common sensing task among UAVs within a more sophisticated \textit{cell-free} network \cite{Bjornson:TC20, DVilor:TWC23, Xu:TC23} that leverages multiple-input multiple-output (MIMO) technology.

In this setup, \textit{multi-antenna} UAVs transmit their sensory data to a cloud server (CS) via \textit{multi-antenna} access points (APs).
These APs are linked to the CS through finite-capacity fronthaul connections, necessitating the use of fronthaul quantization and compression techniques \cite{Zhou:TSP16, Park:TSIPN21, Kim:TVT22}.
Additionally, unlike the approach in \cite{Meng:TWC23}, which solely relied on TDMA transmission for UAVs' private sensory data, we propose a hybrid utilization of TDMA and \textit{non-orthogonal multiple access (NOMA)} \cite{Yang:TC16}. This is motivated by the potential advantages of NOMA over TDMA, especially within cell-free MIMO networks, since the CS can utilize the uplink received signals from all APs to effectively manage the interference signals.

However, the design of the outlined system presents challenges. In fact, one needs to concurrently optimize task allocation between common and per-UAV private sensing tasks. Additionally, one needs to optimize hybrid TDMA-NOMA and cooperative transmission strategies for transmitting sensory data for private and common tasks, respectively.
It is also noted that the choice between employing TDMA and NOMA modes for private sensory data transmission involves a tradeoff depending on system factors such as signal-to-noise ratio (SNR) of the wireless link, capacity of fronthaul links, and network configurations, including the numbers of UAVs, APs, and their antennas.

We establish the problem of jointly optimizing the task allocation variables for sensing tasks, along with hybrid TDMA-NOMA-cooperative transmission strategies, with the aim of minimizing mission completion time.
To tackle the non-convex nature of the problem, we reformulate it using
fractional programming (FP) \cite{Shen:TSP18} and matrix FP techniques \cite{Shen:TN19}.
These techniques were proven effective in optimizing user-cloud assignment and downlink beamforming for multi-cloud radio access networks (RAN) \cite{Ahmad:ICC20}, as well as two-tier receive signal combining for fully-decoupled RAN scenarios \cite{Zhao:TWC23}. Based on the reformulated problem, we propose an iterative algorithm that alternately optimizes primary and auxiliary block variables, resulting in monotonically decreasing completion time values.
Through extensive numerical results, we validate the efficacy of the proposed task allocation and hybrid TDMA-NOMA-cooperative transmission scheme in expediting the completion of sensing mission tasks.
In Table I, we compare this work with recent literature on UAV sensing \cite{Meng:TVT19, Xu:WCSP23,Zhou:IOT24, Liu:TCCN24, Chen:TVT20, Wu:TC20, Meng:TWC23}.

The paper is organized as follows. In Sec. \ref{sec:System-Model}, we describe the system model of a multi-UAV collaborative sensing system, encompassing the wireless channel model from UAVs to APs, the fronthaul connections, and the splitting variables for common and private sensing tasks.
In Sec. \ref{sec:proposed-hybrid}, we outline the overall process of the proposed sensing and hybrid TDMA-NOMA-cooperative transmission system, accompanied by quantification of the total mission completion time.
In Sec. \ref{sec:optimization}, we formulate the optimization problem for the proposed multi-UAV sensing system and develop an iterative optimization algorithm.
In Sec. \ref{sec:numerical-results}, we present numerical results that validate the efficacy of the proposed system.
Finally, the paper concludes in Sec. \ref{sec:conclusion}.

\section{System Model\label{sec:System-Model}}

We consider a collaborative sensing system in which, as illustrated in Fig. \ref{fig:system-model}, $K$ UAVs each equipped with $n_U$ antennas sense an area of interest and transmit the images to a CS which processes the collected sensing information.
Specifically, we assume that the UAVs fly at high altitudes and in close proximity to each other. Under these conditions, it is likely that the UAVs have overlapped fields of view, as illustrated in Fig. \ref{fig:system-model} and supported by previous studies \cite{Meng:TWC23, Xu:TR24, Lin:IOT21}.
Unlike \cite{Meng:TWC23}, which assumed a single receiving BS, we consider a cell-free MIMO network in which the UAVs' transmit signals are cooperatively received by $M$ APs each equipped with $n_A$ antennas and connected to the CS through a dedicated fronthaul link of capacity $C_F$ bits per second (bps).
The bandwidth of the UAVs-to-APs wireless channel is $B$ Hz.
Let us define the sets of UAVs' and APs' indices as $\mathcal{K} = \{1,2,\ldots,K\}$ and $\mathcal{M} = \{1,2,\ldots,M\}$, respectively.

\subsection{Wireless Channel Model} \label{sub:channel-model}

The received signal vector $\mathbf{y}_i\in\mathbb{C}^{n_A\times 1}$ of AP $i$ is given by
\begin{align}
    \mathbf{y}_i = \sum\nolimits_{k\in\mathcal{K}} \mathbf{H}_{i,k}\mathbf{x}_k + \mathbf{z}_i, \label{eq:received-signal-vector}
\end{align}
where $\mathbf{H}_{i,k}\in\mathbb{C}^{n_A\times n_U}$ denotes the channel matrix from UAV $k$ to AP $i$, $\mathbf{x}_k \in \mathbb{C}^{n_U\times 1}$ represents the transmitted signal vector of UAV $k$, and $\mathbf{z}_i\sim \mathcal{CN}(\mathbf{0}, \sigma_z^2\mathbf{I})$ is the additive noise.
Each transmitted signal vector is subject to a power constraint $\mathbb{E}[\|\mathbf{x}_k\|^2] \leq P_U$.
Without loss of generality, we assume that the UAVs' indices are ordered as
\begin{align}
    \|\mathbf{H}_1\|^2_F \leq \|\mathbf{H}_2\|^2_F \leq \ldots \leq \|\mathbf{H}_{K}\|^2_F, \label{eq:ordered-channel-magnitude}
\end{align}
where $\mathbf{H}_k = [\mathbf{H}_{1,k}^H \, \mathbf{H}_{2,k}^H \, \cdots \, \mathbf{H}_{M,k}^H]^H\in\mathbb{C}^{n_A M \times n_U}$ represents the channel matrix from UAV $k$ to all APs.
Also, it is assumed that the CS has a perfect channel state information (CSI) of all matrices $\mathbf{H} = \{\mathbf{H}_{i,k}\}_{i\in\mathcal{M},k\in\mathcal{K}}$ based on which it coordinates the operation of whole system.

\subsection{Sensing Task and Task Splitting} \label{sub:task-partitioning}

We assume that the total sensing task, which needs to be executed by the UAVs, requires a sensing time of $\tau_{\text{total}}^S$ seconds, and
the amount of sensory data representing the sensing result is $b_{\text{total}}$ bits.
We partition the entire sensing task into $K+1$ subtasks, comprising a single common task and $K$ private tasks. The common sensing task is executed by all UAVs, while each private task is exclusively assigned to the corresponding UAV.

Performing the common task with all UAVs might lead to a waste of their individual sensing capabilities. However, as reported in \cite{Meng:TWC23}, it allows for cooperative transmission of the sensing output data among the UAVs, resulting in an improved spectral efficiency of the wireless link.

In contrast, the private tasks are designed to be non-overlapping across UAVs, enabling an efficient use of each UAV's sensing capability. However, since the UAVs need to report independent sensing results, the uplink transmission should address the impact of inter-UAV interference signals.
Given the above discussion, it is essential to optimally choose the task splitting fractions, considering the tradeoff between the sensing and wireless transmission times.

To elaborate, we define the task allocation variables $\boldsymbol{\alpha} = \{\alpha_0, \alpha_1, \ldots, \alpha_K\}$, where $\alpha_0$ and $\alpha_k$ ($k\in\mathcal{K}$) denote the portions of the common task, commonly performed by all UAVs, and the $k$th private task, respectively. Thus, the task allocation variables $\boldsymbol{\alpha}$ satisfy
\begin{align}
    \alpha_k \geq 0, \, \forall k \in \{0\}\cup \mathcal{K} \, \text{ and } \, \alpha_0 + \sum\nolimits_{k\in\mathcal{K}} \alpha_k = 1. \label{eq:constraints-alpha}
\end{align}
Given the ascending order of the channel magnitudes in (\ref{eq:ordered-channel-magnitude}), we allocate a larger portion of the private sensing task to UAV $k$ compared to UAV $k-1$ by imposing $\alpha_{k} \geq \alpha_{k-1}$, $\forall k\in\mathcal{K}\setminus \{1\}$.

\section{Proposed Hybrid TDMA-NOMA-Cooperative Transmission} \label{sec:proposed-hybrid}

\begin{figure}
\centering\includegraphics[width=1.0\linewidth]{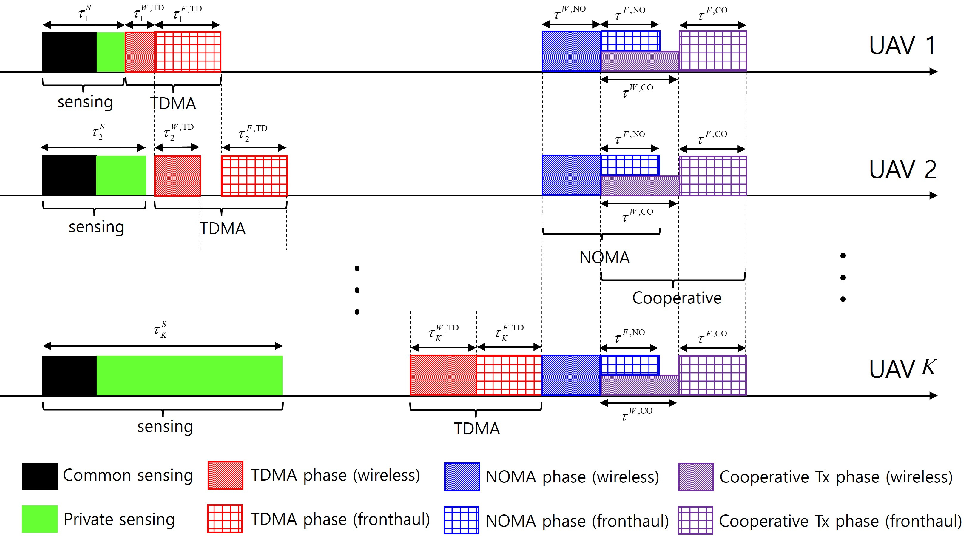}
\vspace{-3mm}
\caption{\small Illustration of the overall process of the proposed collaborative sensing system.} \label{fig:overall-process}
\end{figure}

To mitigate the inter-UAV interference during the transmission of the private sensing results, only TDMA was considered in \cite{Meng:TWC23}.
In this section, we propose an enhanced transmission strategy which combines both TDMA and NOMA transmission modes.
Fig. \ref{fig:overall-process} shows the overall process of the proposed collaborative sensing and communication system.

\subsection{Overall Process of Sensing and Transmission} \label{sub:overall-process}

The process starts with the sensing at UAVs followed by the transmission of the sensory data from UAVs to CS through APs.
Each UAV $k$ executes the common and its private sensing tasks. Thus, the sensing time $\tau_k^S$ of UAV $k$ is given by
\begin{align}
    \tau_k^S = \left(\alpha_0 + \alpha_k\right)\tau_{\text{total}}^S. \label{eq:sensing-time-UAV-k}
\end{align}
Since we imposed $\alpha_{k} \geq \alpha_{k-1}$, $\forall k\in\mathcal{K}\setminus \{1\}$ on the task allocation variables, the sensing time in (\ref{eq:sensing-time-UAV-k}) satisfies the conditions
\begin{align}
    \tau_k^S \geq \tau_{k-1}^S, \, \forall k\in\mathcal{K}\setminus\{1\}. \label{eq:sensing-time-ordered}
\end{align}
Due to the sequential ordering of the sensing time in (\ref{eq:sensing-time-ordered}), the sensing missions of UAVs are completed in the order of UAV 1, UAV 2, $\ldots$, UAV $K$.

Once the sensing tasks are finished, the UAVs are required to transmit the obtained sensory data to the CS. As shown in Fig. \ref{fig:overall-process}, it is assumed that the UAVs first transmit their private sensing results.
Since the private sensory data is independent across UAVs, concurrent transmissions among UAVs cause inter-UAV interference. To avoid interference, reference \cite{Meng:TWC23} considered TDMA transmission, where each UAV is allocated only a portion of the time resource. However, when there are many multi-antenna APs with sufficiently large fronthaul capacity, the NOMA scheme, in which all UAVs utilize the entire time resource, can outperform TDMA, as the inter-UAV interference signals can be efficiently managed by joint decoding at the CS.
To transmit the independent private sensory data while harnessing the complementary benefits of both TDMA and NOMA, the proposed scheme divides the private sensory data of each UAV $k$ into two submessages. These split submessages are sequentially transmitted utilizing the TDMA and NOMA modes.
We define the portions of the submessages communicated in TDMA and NOMA modes as $\alpha_k^{\text{TD}}$ and $\alpha_k^{\text{NO}}$, respectively, which satisfy $\alpha_k^{\text{TD}} \geq 0$, $\alpha_k^{\text{NO}} \geq 0$, and $\alpha_k^{\text{TD}} +  \alpha_k^{\text{NO}} = \alpha_k$.
Therefore, the constraint (\ref{eq:constraints-alpha}) can be rewritten as
\begin{subequations} \label{eq:constraints-alpha-rewritten}
\begin{align}
    &\alpha_0 \geq 0, \, \alpha_k^X \geq 0, \, k\in\mathcal{K}, X\in\{\text{TD}, \text{NO}\}, \label{eq:constraints-alpha-rewritten-1} \\  \text{and} \,\,\, &\alpha_0 + \sum\nolimits_{k\in\mathcal{K}} \left(\alpha_k^{\text{TD}} + \alpha_k^{\text{NO}}\right) = 1. \label{eq:constraints-alpha-rewritten-2}
\end{align}
\end{subequations}
After the completion of transmitting the private sensing data, in the last stage, the UAVs \textit{collaboratively} transmit the results of the common task.
For the NOMA transmission of the split private submessages and the cooperative transmission of the common sensory data, we assume a synchronous transmission among the UAVs.

\subsection{Total Mission Completion Time} \label{sub:completion-time}

To quantify the completion time, we define the time spent by UAV $k$ to transmit its private sensory data in the TDMA mode as $\tau_k^{W, \text{TD}}$.
The superscript $W$ stands for the UAVs-to-APs \textit{wireless access} link which is distinguished from the APs-to-CS \textit{fronthaul} link.
If we define $\tau^{S+W,\text{TD}}_k$ as the time required for UAV $k$ to complete its sensing tasks and its TDMA transmission, it is computed as
\begin{align}
    \tau^{S+W,\text{TD}}_k = \max \big\{ \tau_k^S, \tau^{S+W,\text{TD}}_{k-1} \big\} + \tau_k^{W, \text{TD}}, \label{eq:time-sensing-plus-TDMA-each-UAV}
\end{align}
for all $k\in\mathcal{K}$ with setting $\tau_0^{S+W,\text{TD}} = 0$. The first-term in the right-hand side (RHS) of (\ref{eq:time-sensing-plus-TDMA-each-UAV}) comes from the fact that UAV $k$ can start its TDMA transmission only after both its sensing task and the TDMA transmission of UAV $k-1$ are finished.

Similarly to (\ref{eq:time-sensing-plus-TDMA-each-UAV}), if we denote the time required for the sensing and TDMA transmission of UAV $k$ and the associated fronthaul transmission by $\tau_k^{S+W+F, \text{TD}}$, it is given as
\begin{align}
    \tau_k^{S+W+F, \text{TD}} = \max \big\{ \tau_k^{S+W,\text{TD}}, \tau_{k-1}^{S+W+F, \text{TD}} \big\} + \tau_k^{F, \text{TD}}, \label{eq:time-sensing-plus-TDMA-plus-FH-each-UAV}
\end{align}
where $\tau_k^{F, \text{TD}}$ represents the fronthaul transmission time for delivering the uplink signals received during the TDMA transmission of UAV $k$, and we set $\tau_0^{S+W+F, \text{TD}} = 0$.
Again, the first-term in the RHS of (\ref{eq:time-sensing-plus-TDMA-plus-FH-each-UAV}) indicates that the fronthaul transmission for UAV $k$ begins only after both the TDMA transmission of UAV $k$ and the fronthaul delivery for UAV $k-1$ are finished.

The total completion time of the collaborative sensing and communications from UAVs to CS is given as
\begin{align}
    \tau_{\text{total}} = \tau_K^{S+W+F,\text{TD}} + \tau^{W,\text{NO}} + \max\big\{
    \tau^{F, \text{NO}}, \tau^{W, \text{CO}} \big\} + \tau^{F, \text{CO}}, \label{eq:total-completion-time}
\end{align}
where $\tau^{W,\text{NO}}$ and $\tau^{F,\text{NO}}$ quantify the transmission time on the wireless and fronthaul links, respectively, in the NOMA stage. Similarly, the wireless and fronthaul delays in the cooperative transmission stage are respectively denoted by $\tau^{W,\text{CO}}$ and $\tau^{F,\text{CO}}$.
In subsequent subsections, we discuss the TDMA, NOMA and cooperative transmission phases quantifying the completion time of each phase.

\subsection{TDMA Transmission Phase} \label{sub:TDMA}

In the TDMA phase, each UAV is exclusively assigned a time slot for an interference-free uplink transmission.
On the time resource assigned to UAV $k$, the received signal vector of AP $i$ in (\ref{eq:received-signal-vector}) reduces to
\begin{align}
    \mathbf{y}_i^{\text{TD}} = \mathbf{H}_{i,k} \mathbf{x}_k^{\text{TD}} + \mathbf{z}_i^{\text{TD}}. \label{eq:received-signal-vector-TDMA}
\end{align}
We assume a Gaussian channel coding, i.e., $\mathbf{x}_k^{\text{TD}}\sim\mathcal{CN}(\mathbf{0}, \mathbf{S}_k^{\text{TD}})$ with a covariance matrix $\mathbf{S}_k^{\text{TD}} \succeq \mathbf{0}$ satisfying $\text{tr}(\mathbf{S}_k^{\text{TD}}) \leq P_U$.
Through the fronthaul link of a finite capacity $C_F$ bps, AP $i$ informs the CS of a quantized version of the received signal vector $\mathbf{y}_i^{\text{TD}}$.
Under the assumption of a Gaussian codebook for the fronthaul quantization \cite{Zhou:TSP16,Park:TSIPN21,Kim:TVT22}, the quantized signal $\hat{\mathbf{y}}_i^{\text{TD}}$ recovered by CS can be written as
\begin{align}
    \hat{\mathbf{y}}_i^{\text{TD}} = \mathbf{y}_i^{\text{TD}} + \mathbf{q}_i^{\text{TD}}, \label{eq:quantization}
\end{align}
with the quantization noise vector $\mathbf{q}_i^{\text{TD}}$ distributed as $\mathbf{q}_i^{\text{TD}} \sim \mathcal{CN}(\mathbf{0}, \boldsymbol{\Omega}_{i,k}^{\text{TD}})$.
Following the rate-distortion theoretic framework \cite{Gamal:Cambridge11}, the number of bits required to represent each sample of the signal $\hat{\mathbf{y}}_i$ is given as
\begin{align}
    g_{i,k}^{\text{TD}}\left(\mathbf{S}^{\text{TD}}, \boldsymbol{\Omega}_{i,k}^{\text{TD}}\right) &= I\left(\mathbf{y}_i^{\text{TD}} ; \hat{\mathbf{y}}_i^{\text{TD}}\right) \label{eq:compression-bit-TDMA} \\
    & = \log_2\det\left( \boldsymbol{\Omega}_{i,k}^{\text{TD}} + \sigma_z^2\mathbf{I} +\mathbf{H}_{i,k} \mathbf{S}_k^{\text{TD}}\mathbf{H}_{i,k}^H \right) - \log_2\det\left(\boldsymbol{\Omega}_{i,k}^{\text{TD}}\right), \nonumber
\end{align}
with $\mathbf{S}^{\text{TD}} = \{\mathbf{S}_k^{\text{TD}}\}_{k\in\mathcal{K}}$.
As the time duration of each baseband sample is approximately given as $1/B$ sec, the amount of information that AP $i$ needs to send to the CS on fronthaul is given by $B\cdot\tau_k^{W,\text{TD}} g_{i,k}^{\text{TD}}(\mathbf{S}^{\text{TD}}, \boldsymbol{\Omega}_{i,k}^{\text{TD}})$ bits.
Consequently, the transmission time $\tau_k^{F, \text{TD}}$ on the fronthaul link regarding the TDMA stage of UAV $k$ is given as
\begin{align}
    \tau_k^{F, \text{TD}} = \max_{i\in\mathcal{M}} \frac{B\cdot\tau_k^{W,\text{TD}} g_{i,k}^{\text{TD}}\left(\mathbf{S}^{\text{TD}}, \boldsymbol{\Omega}_{i,k}^{\text{TD}}\right) }{C_F}. \label{eq:time-TDMA-fronthaul}
\end{align}

Based on the quantized signals $\hat{\mathbf{y}}^{\text{TD}} = [\hat{\mathbf{y}}_1^{\text{TD}H} \, \cdots \, \hat{\mathbf{y}}_M^{\text{TD}H}]^H$ available, the CS decodes the signal $\mathbf{x}_k^{\text{TD}}$.
Thus, the transmission rate $R_k^{\text{TD}}$ on the wireless link is given as
\begin{align}
    R_k^{\text{TD}}\! = \! B \log_2\det\left( \mathbf{I} + \left( \sigma_z^2\mathbf{I} + \bar{\boldsymbol{\Omega}}_k^{\text{TD}} \right)^{-1} \mathbf{H}_k \mathbf{S}_k^{\text{TD}} \mathbf{H}_k^H \right), \label{eq:data-rate-TDMA}
\end{align}
where $\bar{\boldsymbol{\Omega}}_k^{\text{TD}} = \text{blkdiag}(\{\boldsymbol{\Omega}_{i,k}^{\text{TD}}\}_{i\in\mathcal{M}})$.
Consequently, the transmission time $\tau_k^{W, \text{TD}}$ on wireless link in the TDMA phase is given as
\begin{align}
    \tau_k^{W, \text{TD}} = \frac{ b_{\text{total}} \alpha_k^{\text{TD}} }{  R_k^{\text{TD}} }. \label{eq:time-TDMA-wireless}
\end{align}

\subsection{NOMA Transmission Phase} \label{sub:NOMA-transmission}

In the NOMA transmission phase, all UAVs concurrently transmit the signals $\mathbf{x}_k^{\text{NO}} \sim \mathcal{CN}(\mathbf{0}, \mathbf{S}_k^{\text{NO}})$, $k\in\mathcal{K}$, where $\mathbf{S}_k^{\text{NO}}\succeq \mathbf{0}$ and $\text{tr}(\mathbf{S}_k^{\text{NO}}) \leq P_U$.
Since the transmitted signals of all UAVs arrive at each AP $i$, the number of bits needed to represent each sample of a quantized version $\hat{\mathbf{y}}_i^{\text{NO}} = \mathbf{y}_i^{\text{NO}} + \mathbf{q}_i^{\text{NO}}$ with $\mathbf{q}_i^{\text{NO}}\sim\mathcal{CN}(\mathbf{0},\boldsymbol{\Omega}_i^{\text{NO}})$ is given by
\begin{align}
    g_i^{\text{NO}}\left( \mathbf{S}^{\text{NO}}, \boldsymbol{\Omega}_i^{\text{NO}} \right)
    =\! \log_2\det\!\left(\! \boldsymbol{\Omega}_i^{\text{NO}}\! +\!  \sigma_z^2\mathbf{I +}\!\!\sum_{k\in\mathcal{K}}\mathbf{H}_{i,k}\mathbf{S}_k^{\text{NO}}\mathbf{H}_{i,k}^H \!\right)\! -\!\log_2\det\!\left(\boldsymbol{\Omega}_i^{\text{NO}}\right), \label{eq:compression-bit-NOMA}
\end{align}
where we have defined the notation $\mathbf{S}^{\text{NO}} = \{\mathbf{S}_k^{\text{NO}}\}_{k\in\mathcal{K}}$.
Accordingly, the transmission time on the fronthaul link is given as
\begin{align}
    \tau^{F,\text{NO}} = \max_{i\in\mathcal{M}} \frac{ B\cdot\tau^{W,\text{NO}} g_i^{\text{NO}}\left( \mathbf{S}^{\text{NO}}, \boldsymbol{\Omega}_i^{\text{NO}} \right) }{C_F}. \label{eq:time-NOMA-fronthaul}
\end{align}

The total quantized signal vector $\hat{\mathbf{y}}^{\text{NO}} = [\hat{\mathbf{y}}^{\text{NO}H}_1 \cdots \hat{\mathbf{y}}^{\text{NO}H}_{N_A}]^H$ received at the CS can be rewritten as
\begin{align}
    \hat{\mathbf{y}}^{\text{NO}} = \sum\nolimits_{l\in\mathcal{K}} \mathbf{H}_l\mathbf{x}_l^{\text{NO}} + \mathbf{z}^{\text{NO}} + \mathbf{q}^{\text{NO}}, \label{eq:total-quantized-signal}
\end{align}
where $\mathbf{q}^{\text{NO}}=[\mathbf{q}^{\text{NO}H}_1\cdots \mathbf{q}^{\text{NO}H}_{N_A}]^H \sim \mathcal{CN}(\mathbf{0}, \bar{\boldsymbol{\Omega}}^{\text{NO}})$ with $\bar{\boldsymbol{\Omega}}^{\text{NO}} = \text{blkdiag}(\{\boldsymbol{\Omega}_i^{\text{NO}}\}_{i\in\mathcal{M}})$.
To decode the signals $\mathbf{x}_1^{\text{NO}},\ldots,\mathbf{x}_K^{\text{NO}}$, the CS adopts a successive interference cancellation (SIC) decoding, whereby the signals are successively decoded while cancelling the impact of the decoded signals from $\hat{\mathbf{y}}^{\text{NO}}$.
To elaborate, we define a permutation $\pi:\mathcal{K}\rightarrow \mathcal{K}$, which defines the SIC decoding order. That is, the signals are decoded in the order of $\mathbf{x}_{\pi(1)}^{\text{NO}}$, $\mathbf{x}_{\pi(2)}^{\text{NO}}$, \ldots, $\mathbf{x}_{\pi(K)}^{\text{NO}}$.
Therefore, the achievable data rate $R_{\pi(k)}^{\text{NO}}$ of $\mathbf{x}_{\pi(k)}^{\text{NO}}$ on the wireless link is thus given as
\begin{align}
    R_{\pi(k)}^{\text{NO}} \!= \!B \log_2\!\det\left( \! \mathbf{I}\! + \!\left( \! \mathbf{N}_{\pi(k)}^{\text{NO}} \! \right)^{\!-1} \mathbf{H}_{\pi(k)} \mathbf{S}_{\pi(k)}^{\text{NO}} \mathbf{H}_{\pi(k)}^H \! \right), \label{eq:data-rate-NOMA}
\end{align}
where we have defined the covariance matrix $\mathbf{N}_{\pi(k)}^{\text{NO}}$ of the interference-plus-noise signals when decoding the signal $\mathbf{x}_{\pi(k)}$ in the NOMA phase as
\begin{align}
    \mathbf{N}_{\pi(k)}^{\text{NO}} = \sum\nolimits_{l=k+1}^K \mathbf{H}_{\pi(l)} \mathbf{S}_{\pi(l)}^{\text{NO}} \mathbf{H}_{\pi(l)}^H + \sigma_z^2\mathbf{I} + \bar{\boldsymbol{\Omega}}^{\text{NO}}. \label{eq:covariance-IF-noise-NOMA}
\end{align}
Consequently, the transmission time $\tau^{W,\text{NO}}$ on the wireless link in the NOMA phase is given by
\begin{align}
    \tau^{W,\text{NO}} = \max_{k\in\mathcal{K}} \, \frac{ b_{\text{total}}\alpha_k^{\text{NO}} }{R_k^{\text{NO}}}. \label{eq:time-NOMA-wireless}
\end{align}

\subsection{Cooperative Transmission Phase} \label{sub:cooperative-transmission}

Since all UAVs execute the common sensing task, the common sensory data is available at all UAVs. Therefore, cooperative transmission across the UAVs can be performed without inter-UAV interference.
To elaborate, we set $\mathbf{x}_k^{\text{CO}} = \mathbf{V}_{k}\mathbf{s}^{\text{CO}}$, where $\mathbf{s}^{\text{CO}}\in\mathbb{C}^{n_S\times 1}$ represents the data symbol vector with $\mathbf{s}^{\text{CO}}\sim\mathcal{CN}(\mathbf{0}, \mathbf{I})$, and $\mathbf{V}_k\in\mathbb{C}^{n_U\times n_S}$ is the precoding matrix that satisfies $\text{tr}(\mathbf{V}_k\mathbf{V}_k^H)\leq P_U$. Here $n_S$ denotes the number of data streams that are simultaneously transmitted.
It is emphasized that the data vector $\mathbf{s}^{\text{CO}}$ is shared by all UAVs, since it is the output of the common sensing task executed by all UAVs.
The received signal vector of each AP $i$ is then given by
\begin{align}
    \mathbf{y}_i^{\text{CO}}\! = \!\left(\sum\nolimits_{k\in\mathcal{K}} \mathbf{H}_{i,k}\mathbf{V}_k\right) \mathbf{s}^{\text{CO}}\! + \mathbf{z}_i^{\text{CO}} = \bar{\mathbf{H}}_i\bar{\mathbf{V}}\mathbf{s}^{\text{CO}}\! + \mathbf{z}_i^{\text{CO}}, \label{eq:received-signal-AP-i-cooperative-tx}
\end{align}
where the matrices $\bar{\mathbf{H}}_i\in\mathbb{C}^{n_A\times n_U K}$ and $\bar{\mathbf{V}}\in\mathbb{C}^{n_U K \times n_S}$ are defined as $\bar{\mathbf{H}}_i = [\mathbf{H}_{i,1} \, \cdots \, \mathbf{H}_{i,K}]$ and $\bar{\mathbf{V}} = [\mathbf{V}_1^H \, \cdots \, \mathbf{V}_K^H]^H$, respectively.
Representing the quantized signal $\hat{\mathbf{y}}_i^{\text{CO}}$ which is transferred to the CS through fronthaul link as $\hat{\mathbf{y}}_i^{\text{CO}} = \mathbf{y}_i^{\text{CO}} + \mathbf{q}_i^{\text{CO}}$ with $\mathbf{q}_i^{\text{CO}}\sim\mathcal{CN}(\mathbf{0},\boldsymbol{\Omega}_i^{\text{CO}})$,
the transmission time $\tau^{F,\text{CO}}$ on the fronthaul link is written as
\begin{align}
    \tau^{F,\text{CO}} = \max_{i\in\mathcal{M}} \frac{ B\cdot\tau^{W,\text{CO}} g_i^{\text{CO}}\left( \mathbf{S}^{\text{CO}}, \boldsymbol{\Omega}_i^{\text{CO}} \right) }{ C_F }, \label{eq:time-cooperative-fronthaul}
\end{align}
where we have defined the notation $\mathbf{S}^{\text{CO}} = \bar{\mathbf{V}}\bar{\mathbf{V}}^H \succeq \mathbf{0}$ and the function $g_i^{\text{CO}}( \mathbf{S}^{\text{CO}}, \boldsymbol{\Omega}_i^{\text{CO}} )$ as
\begin{align}
    g_i^{\text{CO}}\left( \mathbf{S}^{\text{CO}}, \boldsymbol{\Omega}_i^{\text{CO}} \right) &= I\left(\mathbf{y}_i^{\text{CO}}; \hat{\mathbf{y}}_i^{\text{CO}}\right) \label{eq:compression-rate-cooperative}\\
    &= \log_2\det\left( \boldsymbol{\Omega}_i^{\text{CO}} +
 \sigma_z^2\mathbf{I} + \bar{\mathbf{H}}_i\mathbf{S}^{\text{CO}}\bar{\mathbf{H}}_i^H  \right) - \log_2\det\left(\boldsymbol{\Omega}_i^{\text{CO}}\right). \nonumber
\end{align}

The total quantized signal vector $\hat{\mathbf{y}}^{\text{CO}}$ in the cooperative transmission phase is given by
\begin{align}
    \hat{\mathbf{y}}^{\text{CO}} = \bar{\mathbf{H}} \bar{\mathbf{V}}\mathbf{s}^{\text{CO}} + \mathbf{z}^{\text{CO}} + \mathbf{q}^{\text{CO}}, \label{eq:total-quantized-signal-cooperative-tx}
\end{align}
where $\bar{\mathbf{H}} = [\bar{\mathbf{H}}_1^H \, \cdots \, \bar{\mathbf{H}}_M^H]^H\in\mathbb{C}^{n_A M \times n_U K}$ and $\mathbf{q}^{\text{CO}}\sim\mathcal{CN}(\mathbf{0}, \bar{\boldsymbol{\Omega}}^{\text{CO}})$ with $\bar{\boldsymbol{\Omega}}^{\text{CO}} = \text{blkdiag}(\{\boldsymbol{\Omega}_i^{\text{CO}}\}_{i\in\mathcal{M}})$.
From (\ref{eq:total-quantized-signal-cooperative-tx}), the achievable data rate on the wireless link of the cooperative transmission phase is given as
\begin{align}
    R^{\text{CO}} = B \log_2\det\left( \mathbf{I} + \left(
\sigma_z^2\mathbf{I} + \bar{\boldsymbol{\Omega}}^{\text{CO}} \right)^{-1}\bar{\mathbf{H}}\mathbf{S}^{\text{CO}}\bar{\mathbf{H}}^H \right). \label{eq:data-rate-cooperative-tx}
\end{align}
Therefore, the transmission time $\tau^{W,\text{CO}}$ on the wireless link becomes
\begin{align}
    \tau^{W,\text{CO}} = \frac{ b_{\text{total}}\alpha_0 }{R^{\text{CO}}}. \label{eq:time-wireless-cooperative-tx}
\end{align}

\section{Proposed Design Algorithm} \label{sec:optimization}

In this section, we address the joint optimization of the task allocation $\boldsymbol{\alpha} = \{\alpha_0\}\cup \{\alpha_k^{\text{TD}}, \alpha_k^{\text{NO}}\}_{k\in\mathcal{K}}$ and the hybrid TDMA-NOMA-cooperative transmission strategies $\mathbf{S} = \mathbf{S}^{\text{TD}}\cup\mathbf{S}^{\text{NO}} \cup \mathbf{S}^{\text{CO}}$, $\boldsymbol{\Omega} = \{\boldsymbol{\Omega}_{i,k}^{\text{TD}}\}_{i\in\mathcal{M},k\in\mathcal{K}}\cup \{\boldsymbol{\Omega}_i^{\text{NO}}, \boldsymbol{\Omega}_i^{\text{CO}}\}_{i\in\mathcal{M}}$ with the goal of minimizing the mission completion time $\tau_{\text{total}}$ in (\ref{eq:total-completion-time}).
We formulate the problem as
\begingroup
\allowdisplaybreaks
\begin{subequations} \label{eq:problem-original}
\begin{align}
    \!\!\!\!\!\underset{\boldsymbol{\alpha}, \mathbf{S}, \boldsymbol{\Omega}, \boldsymbol{\tau}, \mathbf{R}} {\mathrm{min.}}\,\,\,
    & \tau_K^{S+W+F,\text{TD}} \!+\! \tau^{W,\text{NO}} \!+\! \max\big\{ \tau^{F, \text{NO}}\!, \tau^{W, \text{CO}} \big\} \!+\! \tau^{F, \text{CO}} \label{eq:problem-original-cost} \\
    \mathrm{s.t. }\,\,\,\,\,\,\,\,\,\,
    & \tau_k^{F,\text{TD}} \geq \frac{B\cdot\tau_k^{W,\text{TD}} g_{i,k}^{\text{TD}}\left(\mathbf{S}^{\text{TD}}, \boldsymbol{\Omega}_{i,k}^{\text{TD}}\right) }{C_F}, i\in\mathcal{M}, \,k\in\mathcal{K}, \label{eq:problem-original-time-fronthaul-TDMA} \\
    & \tau_k^{W,\text{TD}} \geq \frac{b_{\text{total}}\alpha_k^{\text{TD}}}{R_k^{\text{TD}}}, \, k\in\mathcal{K}, \label{eq:problem-original-time-wireless-TD} \\
    & R_k^{\text{TD}} \leq   B \! \log_2\det\left( \! \mathbf{I} \!+\! \Big( \sigma_z^2\mathbf{I} \!+\!\bar{\boldsymbol{\Omega}}_k^{\text{TD}} \Big)^{\!-1} \mathbf{H}_k \mathbf{S}_k^{\text{TD}} \mathbf{H}_k^H
    \! \right)\!, \nonumber \\
    &\qquad\quad k\in\mathcal{K} \label{eq:problem-original-rate-TDMA} \\
    & \tau_k^S = \left(\alpha_0 + \alpha_k^{\text{TD}} + \alpha_k^{\text{NO}}\right)\tau^S_{\text{total}}, \, k\in\mathcal{K}, \label{eq:problem-original-time-sensing-1} \\
    & \tau_k^{S+W,\text{TD}} \geq \max\{\tau_k^S, \tau_{k-1}^{S+W,\text{TD}}\} + \tau_k^{W,\text{TD}}, \, k\in\mathcal{K}, \label{eq:problem-original-time-sensing-wireless-TDMA} \\
    & \tau_k^{S+W+F, \text{TD}} \geq \label{eq:problem-original-time-sensing-wireless-fronthaul-TDMA} \\
    & \! \max \big\{ \tau_k^{S+W,\text{TD}}\!, \tau_{k-1}^{S+W+F, \text{TD}} \big\} + \tau_k^{F, \text{TD}}\!, \, k\in\mathcal{K}, \nonumber \\
    & \tau^{F,\text{NO}} \geq  \frac{ B\cdot\tau^{W,\text{NO}} g_i^{\text{NO}}\left( \mathbf{S}^{\text{NO}}, \boldsymbol{\Omega}_i^{\text{NO}} \right) }{C_F}, \, i\in\mathcal{M}, \label{eq:problem-original-time-fronthaul-NOMA} \\
    & \tau^{W,\text{NO}} \geq \frac{b_{\text{total}}\alpha_k^{\text{NO}}}{R_k^{\text{NO}}}, \, k\in\mathcal{K}, \label{eq:problem-original-time-wireless-NOMA} \\
    & R_{\pi(k)}^{\text{NO}} \label{eq:problem-original-rate-wireless-NOMA} \leq \\ & B \! \log_2 \! \det \! \left(\! \mathbf{I} \!+\! \left(\! \mathbf{N}_{\pi(k)}^{\text{NO}} \!\right)^{\!-1} \!\mathbf{H}_{\pi(k)} \mathbf{S}_{\pi(k)}^{\text{NO}} \mathbf{H}_{\pi(k)}^H \!
    \right)\!\!, k\in\mathcal{K} ,\nonumber \\
    & \tau^{F,\text{CO}} \geq \frac{ B\cdot\tau^{W,\text{CO}} g_i^{\text{CO}}\left( \mathbf{S}^{\text{CO}}, \boldsymbol{\Omega}_i^{\text{CO}} \right) }{ C_F }, \, i\in\mathcal{M}, \label{eq:problem-original-time-fronthaul-cooperative} \\
    & \tau^{W,\text{CO}} \geq \frac{b_{\text{total}}\alpha_0}{R^{\text{CO}}}, \label{eq:problem-original-time-wireless-cooperative} \\
    & R^{\text{CO}} \leq  B \! \log_2 \! \det \! \left( \! \mathbf{I} \!+\! \left( \sigma_z^2\mathbf{I} \!+\! \bar{\boldsymbol{\Omega}}^{\text{CO}} \right)^{\!-1} \! \bar{\mathbf{H}}\mathbf{S}^{\text{CO}}\bar{\mathbf{H}}^H
    \! \right)\!, \label{eq:problem-original-rate-wireless-cooperative} \\
    & \text{tr}(\mathbf{S}_k^{\text{TD}}) \leq P_U, \, k\in\mathcal{K}, \label{eq:problem-original-TDMA-power-constraint} \\
    & \text{tr}(\mathbf{S}_k^{\text{NO}}) \leq P_U, \, k\in\mathcal{K}, \label{eq:problem-original-NOMA-power-constraint} \\
    & \text{tr}( \mathbf{E}_k^H\mathbf{S}^{\text{CO}}\mathbf{E}_k ) \leq P_U, \, k\in\mathcal{K}, \label{eq:problem-original-cooperative-power-constraint} \\
    & \alpha_0 + \sum\nolimits_{k\in\mathcal{K}} \left(\alpha_k^{\text{TD}} + \alpha_k^{\text{NO}}\right) = 1, \label{eq:problem-original-alpha-1} \\
    & \alpha_k\geq\alpha_{k-1}, \, k\in\mathcal{K}\setminus\{1\}, \label{eq:problem-original-alpha-2}
\end{align}
\end{subequations}
\endgroup
where $\boldsymbol{\tau} = \{\tau_k^S,\tau_k^{W,\text{TD}},\tau_k^{F,\text{TD}}, \tau_k^{S+W,\text{TD}}, \tau_k^{S+W+F,\text{TD}}\}_{k\in\mathcal{K}}\cup\{\tau^{W,\text{NO}},\tau^{F,\text{NO}},\tau^{W,\text{CO}}, \tau^{F,\text{CO}}\}$ and $\mathbf{R} = \{R_k^{\text{TD}}, R_k^{\text{NO}}\}_{k\in\mathcal{K}}\cup \{R^{\text{CO}}\}$.
We have also defined a shaping matrix $\mathbf{E}_k \in \mathbb{C}^{K n_U \times n_U}$ as
\begin{align}
    \mathbf{E}_k = \left[ \mathbf{0}_{n_U \times (k-1)n_U} \,\, \mathbf{I}_{n_U} \,\, \mathbf{0}_{n_U \times (K-k)n_U} \right]^H.
\end{align}
Our focus is on optimizing $\{\boldsymbol{\alpha}, \mathbf{S}, \boldsymbol{\Omega}\}$ for a fixed SIC decoding order $\pi$ during the NOMA phase, with the optimization of $\pi$ left for future study.
We also remark that including the placements of UAVs in the design space would provide additional gains \cite{Li:JCIN18, Hammouti:TWC19, Guo:TVT20}. However, the UAVs' positions affect the channel matrices in a complex manner, since it influences both the pathloss and the LoS components. This dependence of the channel matrices, and hence the data rates, on the UAVs' placements complicates the optimization of UAVs' placements. Additionally, in the presence of non-LoS (NLoS) components, the data rates are random functions of the position vectors, asking for the derivation of a closed-form expression for the expected data rates with respect to the NLoS components.
Given the above two challenges, we leave the optimization of UAVs' placements for future work.

The problem (\ref{eq:problem-original}) is not a convex problem due to the constraints (\ref{eq:problem-original-time-fronthaul-TDMA})-(\ref{eq:problem-original-rate-TDMA}) and (\ref{eq:problem-original-time-fronthaul-NOMA})-(\ref{eq:problem-original-rate-wireless-cooperative}).
In what follows, we discuss how to convexify those constraints while classifying them into three categories: \textit{i)} fronthaul latency constraints (\ref{eq:problem-original-time-fronthaul-TDMA}), (\ref{eq:problem-original-time-fronthaul-NOMA}), and (\ref{eq:problem-original-time-fronthaul-cooperative}); \textit{ii)} wireless latency constraints (\ref{eq:problem-original-time-wireless-TD}), (\ref{eq:problem-original-time-fronthaul-NOMA}), and (\ref{eq:problem-original-time-wireless-cooperative}); and \textit{iii)} data rate constraints (\ref{eq:problem-original-rate-TDMA}), (\ref{eq:problem-original-rate-wireless-NOMA}), and (\ref{eq:problem-original-rate-wireless-cooperative}).

\subsection{Fronthaul Latency Constraints (\ref{eq:problem-original-time-fronthaul-TDMA}), (\ref{eq:problem-original-time-fronthaul-NOMA}), and (\ref{eq:problem-original-time-fronthaul-cooperative})} \label{sub:convexifying-fronthaul-latency}

Among the non-convex fronthaul latency constraints,
we focus on the constraint (\ref{eq:problem-original-time-fronthaul-TDMA}), as the other constraints (\ref{eq:problem-original-time-fronthaul-NOMA}) and (\ref{eq:problem-original-time-fronthaul-cooperative}) can be addressed similarly. Utilizing the quadratic transform for FP \cite[Cor. 1]{Shen:TSP18} and Fenchel's inequality for $\log_2 \det (\cdot)$ \cite[Lem. 1]{Zhou:TSP16}, (\ref{eq:problem-original-time-fronthaul-TDMA}) is satisfied if and only if there exist $\{\theta_k^{F, \text{TD}}\}_{k\in\mathcal{K}}$, $\{G_{i,k}^{\text{TD}}\}_{i\in\mathcal{M}, k\in\mathcal{K}}$, and $\{\boldsymbol{\Sigma}_{i,k}^{\text{TD}}\succeq \mathbf{0}\}_{i\in\mathcal{M}, k\in\mathcal{K}}$ for which the following conditions hold:
\begin{subequations} \label{eq:convexified-time-fronthaul-TDMA}
\begin{align}
    &2\theta_k^{F,\text{TD}} \sqrt{\tau_k^{F,\text{TD}}} - \left(\theta_k^{F,\text{TD}}\right)^2 \tau_k^{W,\text{TD}} \geq \frac{B}{C_F} G_{i,k}^{\text{TD}}, \, i\in\mathcal{M}, k\in\mathcal{K}, \label{eq:convexified-time-fronthaul-TDMA-1} \\
    & G_{i,k}^{\text{TD}} \geq \log_2\det\left(\boldsymbol{\Sigma}_{i,k}^{\text{TD}}\right) \label{eq:convexified-time-fronthaul-TDMA-2} \\
    & \quad + \frac{1}{\ln 2} \text{tr}\left( \left( \boldsymbol{\Sigma}_{i,k}^{\text{TD}}\right)^{-1}  \left( \boldsymbol{\Omega}_{i,k}^{\text{TD}} + \sigma_z^2\mathbf{I} +\mathbf{H}_{i,k} \tilde{\mathbf{S}}_k^{\text{TD}}\tilde{\mathbf{S}}_k^{\text{TD}H}\mathbf{H}_{i,k}^H \right) \right)  \nonumber \\
    & \quad - \frac{n_A}{\ln 2} -\log_2\det\left( \boldsymbol{\Omega}_{i,k}^{\text{TD}} \right), \, i\in\mathcal{M}, k\in\mathcal{K}. \nonumber
\end{align}
\end{subequations}
In (\ref{eq:convexified-time-fronthaul-TDMA}), we have defined $\tilde{\mathbf{S}}_k^{\text{TD}}\in\mathbb{C}^{n_U\times n_U}$ which satisfies $\mathbf{S}_k^{\text{TD}} = \tilde{\mathbf{S}}_k^{\text{TD}}\tilde{\mathbf{S}}_k^{\text{TD}H}$.
The above constraints (\ref{eq:convexified-time-fronthaul-TDMA}) become convex constraints if the auxiliary variables $\{\theta_k^{F, \text{TD}}\}_{ k\in\mathcal{K}}$ and $\{\boldsymbol{\Sigma}_{i,k}^{\text{TD}}\succeq \mathbf{0}\}_{i\in\mathcal{M}, k\in\mathcal{K}}$ are fixed.
Also, the optimal auxiliary variables that make (\ref{eq:convexified-time-fronthaul-TDMA}) equivalent to (\ref{eq:problem-original-time-fronthaul-TDMA}) are given as
\begin{subequations} \label{eq:optimal-auxiliary-time-fronthaul-TDMA}
\begin{align}
    \theta_k^{F,\text{TD}} &= \sqrt{\tau_k^{F,\text{TD}}} / \tau_k^{W,\text{TD}}, \, k\in\mathcal{K}, \label{eq:optimal-theta-TD} \\
    \boldsymbol{\Sigma}_{i,k}^{\text{TD}} &= \boldsymbol{\Omega}_{i,k}^{\text{TD}} + \sigma_z^2\mathbf{I} +\mathbf{H}_{i,k} \tilde{\mathbf{S}}_k^{\text{TD}}\tilde{\mathbf{S}}_k^{\text{TD}H}\mathbf{H}_{i,k}^H, \, i\in\mathcal{M}, k\in\mathcal{K}.\label{eq:optimal-Sigma-TD}
\end{align}
\end{subequations}
In a similar way, we can restate the constraints (\ref{eq:problem-original-time-fronthaul-NOMA}) and (\ref{eq:problem-original-time-fronthaul-cooperative}) as follows:
\begingroup
\allowdisplaybreaks
\begin{subequations} \label{eq:convexified-time-fronthaul-NOMA-cooperative}
\begin{align}
    & 2\theta^{F,\text{NO}} \sqrt{\tau^{F,\text{NO}}} - \left(\theta^{F,\text{NO}}\right)^2 \tau^{W,\text{NO}} \geq  \frac{B}{C_F} G_i^{\text{NO}}\!, \, i\in\mathcal{M}, \label{eq:convexified-time-fronthaul-NOMA-1} \\
    & G_i^{\text{NO}} \geq \log_2\det\left(\boldsymbol{\Sigma}_i^{\text{NO}}\right) \label{eq:convexified-time-fronthaul-NOMA-2} \\
    & \quad + \frac{1}{\ln 2} \text{tr}\left( \left(\boldsymbol{\Sigma}_i^{\text{NO}}\right)^{-1}\left(
    \begin{array}{c}
    \boldsymbol{\Omega}_i^{\text{NO}} +  \sigma_z^2\mathbf{I} \, + \\ \sum\nolimits_{k\in\mathcal{K}}\mathbf{H}_{i,k}\tilde{\mathbf{S}}_k^{\text{NO}}\tilde{\mathbf{S}}_k^{\text{NO}H}\mathbf{H}_{i,k}^H
    \end{array}
    \right) \right) \nonumber \\
    & \quad - \frac{n_A}{\ln 2} -\log_2\det\left(\boldsymbol{\Omega}_i^{\text{NO}}\right), \, {i\in\mathcal{M}}, \nonumber \\
    & 2\theta^{F,\text{CO}}\sqrt{\tau^{F,\text{CO}}} - \left(\theta^{F,\text{CO}}\right)^2 \tau^{W,\text{CO}} \geq \frac{B}{C_F}G_i^{\text{CO}}, \, i\in\mathcal{M}, \label{eq:convexified-time-fronthaul-cooperative-1} \\
    &G_i^{\text{CO}} \geq \log_2\det\left(\boldsymbol{\Sigma}_i^{\text{CO}}\right) \label{eq:convexified-time-fronthaul-cooperative-2} \\
    & \quad + \frac{1}{\ln 2} \text{tr}\left(\left(\boldsymbol{\Sigma}_i^{\text{CO}}\right)^{-1} \left( \boldsymbol{\Omega}_i^{\text{CO}} \!+\!
    \sigma_z^2\mathbf{I} \!+\! \bar{\mathbf{H}}_i\tilde{\mathbf{S}}^{\text{CO}}\tilde{\mathbf{S}}^{\text{CO}H}\bar{\mathbf{H}}_i^H \right) \right) \nonumber \\
    & \quad - \frac{n_A}{\ln 2}  - \log_2\det\left(\boldsymbol{\Omega}_i^{\text{CO}}\right), \, i\in\mathcal{M}, \nonumber
\end{align}
\end{subequations}
\endgroup
where the optimal $\theta^{F,\text{NO}}$, $\{\boldsymbol{\Sigma}_i^{\text{NO}}\}_{i\in\mathcal{M}}$, $\theta^{F,\text{CO}}$, and $\{\boldsymbol{\Sigma}_i^{\text{CO}}\}_{i\in\mathcal{M}}$ are given as
\begin{subequations} \label{eq:optimal-auxiliary-time-fronthaul-NOMA-cooperative}
\begin{align}
    \theta^{F,\text{NO}} &= \sqrt{\tau^{F,\text{NO}}} / \tau^{W,\text{NO}},  \label{eq:optimal-theta-NOMA} \\
    \boldsymbol{\Sigma}_i^{\text{NO}} &= \boldsymbol{\Omega}_i^{\text{NO}} +  \sigma_z^2\mathbf{I + }\sum\nolimits_{k\in\mathcal{K}}\mathbf{H}_{i,k}\tilde{\mathbf{S}}_k^{\text{NO}}\tilde{\mathbf{S}}_k^{\text{NO}H}\mathbf{H}_{i,k}^H,\, i\in\mathcal{M}, \label{eq:optimal-Sigmal-NOMA} \\
    \theta^{F,\text{CO}} &= \sqrt{\tau^{F,\text{CO}}} / \tau^{W,\text{CO}}, \label{eq:optimal-theta-cooperative} \\
    \boldsymbol{\Sigma}_i^{\text{CO}} &= \boldsymbol{\Omega}_i^{\text{CO}} +
 \sigma_z^2\mathbf{I} + \bar{\mathbf{H}}_i\tilde{\mathbf{S}}^{\text{CO}}\tilde{\mathbf{S}}^{\text{CO}H}\bar{\mathbf{H}}_i^H, \, i\in\mathcal{M}. \label{eq:optimal-Sigma-cooperative}
\end{align}
\end{subequations}
In (\ref{eq:convexified-time-fronthaul-NOMA-cooperative}), we have also defined $\tilde{\mathbf{S}}_k^{\text{NO}}\in\mathbb{C}^{n_U\times n_U}$ and $\tilde{\mathbf{S}}^{\text{CO}}\in\mathbb{C}^{Kn_U\times Kn_U}$ which satisfy $\mathbf{S}_k^{\text{NO}} = \tilde{\mathbf{S}}_k^{\text{NO}}\tilde{\mathbf{S}}_k^{\text{NO}H}$ and $\mathbf{S}^{\text{CO}} = \tilde{\mathbf{S}}^{\text{CO}}\tilde{\mathbf{S}}^{\text{CO}H}$, respectively.
The constraints (\ref{eq:convexified-time-fronthaul-NOMA-cooperative})
are convex constraints if the variables $\theta^{F,\text{NO}}$, $\{\boldsymbol{\Sigma}_i^{\text{NO}}\}_{i\in\mathcal{M}}$, $\theta^{F,\text{CO}}$, and $\{\boldsymbol{\Sigma}_i^{\text{CO}}\}_{i\in\mathcal{M}}$ are fixed.

\subsection{Wireless Latency Constraints (\ref{eq:problem-original-time-wireless-TD}), (\ref{eq:problem-original-time-fronthaul-NOMA}), and (\ref{eq:problem-original-time-wireless-cooperative})} \label{sub:convexifying-wireless-latency}

Rewriting (\ref{eq:problem-original-time-wireless-TD}) as $\tau_k^{W,\text{TD}}/\alpha_k^{\text{TD}} \geq b_{\text{total}} / R_k^{\text{TD}}$ and applying \cite[Cor. 1]{Shen:TSP18}, we obtain the following constraint:
\begin{align}
    2\theta_k^{W,\text{TD}}\sqrt{\tau_k^{W,\text{TD}}} - \left(\theta_k^{W,\text{TD}}\right)^2\alpha_k^{\text{TD}} \geq \frac{b_{\text{total}}}{R_k^{\text{TD}}}, \, k\in\mathcal{K}, \label{eq:convexified-time-wireless-TDMA}
\end{align}
which is a convex constraint if $\{\theta_k^{W,\text{TD}}\}_{k\in\mathcal{K}}$ are fixed.
Also, (\ref{eq:convexified-time-wireless-TDMA}) becomes an equivalent constraint to (\ref{eq:problem-original-time-wireless-TD}) when
\begin{align}
    \theta_k^{W,\text{TD}} = \sqrt{\tau_k^{W,\text{TD}}} / \alpha_k^{\text{TD}}, \,k\in\mathcal{K}. \label{eq:optimal-auxiliary-time-wireless-TDMA}
\end{align}
Similarly, we convert the other wireless latency constraints (\ref{eq:problem-original-time-fronthaul-NOMA}) and (\ref{eq:problem-original-time-wireless-cooperative}) into
\begin{subequations} \label{eq:convexified-time-wireless-NOMA-cooperative}
\begin{align}
    & 2\theta_k^{W,\text{NO}}\sqrt{\tau^{W,\text{NO}}} - \left(\theta_k^{W,\text{NO}}\right)^2\alpha_k^{\text{NO}} \geq \frac{b_{\text{total}}}{R_k^{\text{NO}}}, \, k\in\mathcal{K},\label{eq:convexified-time-wireless-NOMA} \\
    & 2 \theta^{W,\text{CO}} \sqrt{\tau^{W,\text{CO}}} - \left(\theta^{W,\text{CO}}\right)^2 \alpha_0 \geq \frac{b_{\text{total}}}{R^{\text{CO}}}, \label{eq:convexified-time-wireless-cooperative}
\end{align}
\end{subequations}
where the optimal $\{\theta_k^{W,\text{NO}}\}_{k\in\mathcal{K}}$ and $\theta^{W,\text{CO}}$ are given as
\begin{subequations} \label{eq:optimal-auxiliary-time-wireless-NOMA-cooperative}
\begin{align}
    \theta_k^{W,\text{NO}} &= \sqrt{\tau^{W,\text{NO}}} / \alpha_k^{\text{NO}}, \, k\in\mathcal{K},\label{eq:optimal-auxiliary-time-wireless-NOMA} \\
    \theta^{W,\text{CO}} &= \sqrt{\tau^{W,\text{CO}}} / \alpha_0.\label{eq:optimal-auxiliary-time-wireless-cooperative}
\end{align}
\end{subequations}

\subsection{Data Rate Constraints (\ref{eq:problem-original-rate-TDMA}), (\ref{eq:problem-original-rate-wireless-NOMA}), and (\ref{eq:problem-original-rate-wireless-cooperative})} \label{sub:convexifying-data-rate}

We adopt the matrix Lagrangian dual transform in \cite[Thm. 2]{Shen:TN19} on the rate constraint (\ref{eq:problem-original-rate-TDMA}) for the TDMA transmission obtaining the following constraint:
\begin{align}
     \frac{1}{B} & R_k^{\text{TD}} \leq \log_2\det\big(\mathbf{I} + \boldsymbol{\Gamma}_k^{\text{TD}}\big) - \frac{1}{\ln 2}\text{tr}\big(\boldsymbol{\Gamma}_k^{\text{TD}}\big) \label{eq:matrix-Lagrange-dual-transform-rate-TDMA} \\
     & + \frac{1}{\ln 2} \text{tr}\Big[ \big(\mathbf{I} + \boldsymbol{\Gamma}_k^{\text{TD}}\big) \big( 2 \tilde{\mathbf{S}}_k^{\text{TD} H} \mathbf{H}_k^H \boldsymbol{\Phi}_k^{\text{TD}} \nonumber \\
     & - \boldsymbol{\Phi}_k^{\text{TD}H} \big( \sigma_z^2\mathbf{I}
     + \bar{\boldsymbol{\Omega}}_k^{\text{TD}} + \mathbf{H}_k \tilde{\mathbf{S}}_k^{\text{TD}}\tilde{\mathbf{S}}_k^{\text{TD}H}\mathbf{H}_k^H\big) \boldsymbol{\Phi}_k^{\text{TD}}  \big)\Big], \, k\in\mathcal{K},
     \nonumber
\end{align}
which is stricter than (\ref{eq:problem-original-rate-TDMA}) for any $\{\boldsymbol{\Gamma}_k^{\text{TD}}\in\mathbb{C}^{n_U\times n_U }\}_{k\in\mathcal{K}}$ with $\boldsymbol{\Gamma}_k^{\text{TD}} \succeq \mathbf{0}$ and $\{\boldsymbol{\Phi}_k^{\text{TD}}\in\mathbb{C}^{n_A M \times n_U}\}_{k\in\mathcal{K}}$.
In (\ref{eq:matrix-Lagrange-dual-transform-rate-TDMA}), we have defined $\tilde{\mathbf{S}}_k^{\text{TD}}\in\mathbb{C}^{n_U\times n_U}$ which satisfies $\mathbf{S}_k^{\text{TD}} = \tilde{\mathbf{S}}_k^{\text{TD}}\tilde{\mathbf{S}}_k^{\text{TD}H}$.
We note that the constraint (\ref{eq:matrix-Lagrange-dual-transform-rate-TDMA}) becomes a convex constraint if $\{\boldsymbol{\Gamma}_k^{\text{TD}}\}_{k\in\mathcal{K}}$ and $\{\boldsymbol{\Phi}_k^{\text{TD}}\}_{k\in\mathcal{K}}$ are fixed.
Also, the constraint (\ref{eq:matrix-Lagrange-dual-transform-rate-TDMA}) becomes equivalent to (\ref{eq:problem-original-rate-TDMA}) if $\{\boldsymbol{\Gamma}_k^{\text{TD}}\}_{k\in\mathcal{K}}$ and $\{\boldsymbol{\Phi}_k^{\text{TD}}\}_{k\in\mathcal{K}}$ are equal to
\begin{subequations} \label{eq:optimal-auxiliary-rate-TDMA}
\begin{align}
    \boldsymbol{\Gamma}_k^{\text{TD}} &= \tilde{\mathbf{S}}_k^{\text{TD} H} \mathbf{H}_k^H\left( \sigma_z^2\mathbf{I} + \bar{\boldsymbol{\Omega}}_k^{\text{TD}} \right)^{-1} \mathbf{H}_k\tilde{\mathbf{S}}_k^{\text{TD}}, \, k\in\mathcal{K}, \label{eq:optimal-auxiliary-rate-TDMA-1} \\
    \boldsymbol{\Phi}_k^{\text{TD}} &= \left( \sigma_z^2\mathbf{I} + \bar{\boldsymbol{\Omega}}_k^{\text{TD}} + \mathbf{H}_k \tilde{\mathbf{S}}_k^{\text{TD}}\tilde{\mathbf{S}}_k^{\text{TD}H}\mathbf{H}_k^H \right)^{-1} \mathbf{H}_k\tilde{\mathbf{S}}_k^{\text{TD}}, \, k\in\mathcal{K}.\label{eq:optimal-auxiliary-rate-TDMA-2}
\end{align}
\end{subequations}

The matrix Lagrangian dual transform \cite[Thm. 2]{Shen:TN19} can be similarly applied to the constraints (\ref{eq:problem-original-rate-wireless-NOMA}) and (\ref{eq:problem-original-rate-wireless-cooperative}) obtaining the following constraint:
\begin{subequations} \label{eq:matrix-Lagrange-dual-transform-rate-NOMA-cooperative}
\begin{align}
    \frac{1}{B} & R_{\pi(k)}^{\text{NO}} \leq \log_2\det\big(\mathbf{I} + \boldsymbol{\Gamma}_{\pi(k)}^{\text{NO}}\big) - \frac{1}{\ln 2}\text{tr}\big(\boldsymbol{\Gamma}_{\pi(k)}^{\text{NO}}\big) \label{eq:matrix-Lagrange-dual-transform-rate-NOMA} \\
    & + \frac{1}{\ln 2} \text{tr}\Big[  \big(\mathbf{I} + \boldsymbol{\Gamma}_{\pi(k)}^{\text{NO}}\big)\big( 2\tilde{\mathbf{S}}_{\pi(k)}^{\text{NO}H}\mathbf{H}_{\pi(k)}^H \boldsymbol{\Phi}_{\pi(k)}^{\text{NO}} \nonumber \\
    & -\boldsymbol{\Phi}_{\pi(k)}^{\text{NO}H}\big(\mathbf{N}_{\pi(k)}^{\text{NO}}  \!+\!\mathbf{H}_{\pi(k)}\tilde{\mathbf{S}}_{\pi(k)}^{\text{NO}}\tilde{\mathbf{S}}_{\pi(k)}^{\text{NO}H} \mathbf{H}_{\pi(k)}^H\big)\boldsymbol{\Phi}_{\pi(k)}^{\text{NO}} \big)  \Big], \, k\in\mathcal{K}, \nonumber \\
    \frac{1}{B}&R^{\text{CO}}\leq \log_2\det\big(\mathbf{I} + \boldsymbol{\Gamma}^{\text{CO}}\big) - \frac{1}{\ln 2} \text{tr}\big(\boldsymbol{\Gamma}^{\text{CO}}\big)  \label{eq:matrix-Lagrange-dual-transform-rate-cooperative} \\
    & + \frac{1}{\ln 2} \text{tr}\Big[ \big(\mathbf{I} + \boldsymbol{\Gamma}^{\text{CO}}\big)\big( 2\tilde{\mathbf{S}}^{\text{CO}H}\bar{\mathbf{H}}^H \boldsymbol{\Phi}^{\text{CO}} \nonumber \\
    & - \boldsymbol{\Phi}^{\text{CO}H}\big(\sigma_z^2\mathbf{I}
    + \bar{\boldsymbol{\Omega}}^{\text{CO}} + \bar{\mathbf{H}}\tilde{\mathbf{S}}^{\text{CO}}\tilde{\mathbf{S}}^{\text{CO}H}\bar{\mathbf{H}}^H\big)\boldsymbol{\Phi}^{\text{CO}} \big) \Big], \nonumber
\end{align}
\end{subequations}
where the optimal $\{\boldsymbol{\Gamma}_{\pi(k)}^{\text{NO}}\in\mathbb{C}^{n_U\times n_U}\}_{k\in\mathcal{K}}$, $\{\boldsymbol{\Phi}_{\pi(k)}^{\text{NO}}\in\mathbb{C}^{n_A M\times n_U}\}_{k\in\mathcal{K}}$, $\boldsymbol{\Gamma}^{\text{CO}}\in\mathbb{C}^{n_U K\times n_U K}$, and $\boldsymbol{\Phi}^{\text{CO}}\in\mathbb{C}^{n_A M \times n_U K}$ are given as
\begin{subequations} \label{eq:optimal-auxiliary-rate-NOMA-cooperative}
\begin{align}
    \boldsymbol{\Gamma}_{\pi(k)}^{\text{NO}} = & \,\tilde{\mathbf{S}}_{\pi(k)}^{\text{NO}H} \mathbf{H}_{\pi(k)}^H \left(\mathbf{N}_{\pi(k)}^{\text{NO}}\right)^{-1}\mathbf{H}_{\pi(k)}\tilde{\mathbf{S}}_{\pi(k)}^{\text{NO}},\, k\in\mathcal{K}, \label{eq:optimal-auxiliary-rate-NOMA-1}\\
    \boldsymbol{\Phi}_{\pi(k)}^{\text{NO}}=&\left( \mathbf{N}_{\pi(k)}^{\text{NO}} + \mathbf{H}_{\pi(k)}\tilde{\mathbf{S}}_{\pi(k)}^{\text{NO}}\tilde{\mathbf{S}}_{\pi(k)}^{\text{NO}H}  \mathbf{H}_{\pi(k)}^H\right)^{-1} \mathbf{H}_{\pi(k)}\tilde{\mathbf{S}}_{\pi(k)}^{\text{NO}},\, k\in\mathcal{K}, \label{eq:optimal-auxiliary-rate-NOMA-2}\\
    \boldsymbol{\Gamma}^{\text{CO}} =& \,\tilde{\mathbf{S}}^{\text{CO}H}\bar{\mathbf{H}}^H\left(\sigma_z^2\mathbf{I} + \bar{\boldsymbol{\Omega}}^{\text{CO}}\right)^{-1}\bar{\mathbf{H}}\tilde{\mathbf{S}}^{\text{CO}}, \label{eq:optimal-auxiliary-rate-cooperative-1} \\
    \boldsymbol{\Phi}^{\text{CO}}=& \left(\sigma_z^2\mathbf{I} + \bar{\boldsymbol{\Omega}}^{\text{CO}} + \bar{\mathbf{H}}\tilde{\mathbf{S}}^{\text{CO}}\tilde{\mathbf{S}}^{\text{CO}H}\bar{\mathbf{H}}^H\right)^{-1} \bar{\mathbf{H}}\tilde{\mathbf{S}}^{\text{CO}}. \label{eq:optimal-auxiliary-rate-cooperative-2}
\end{align}
\end{subequations}

\subsection{Proposed Algorithm} \label{sub:optimization-algorithm}

Based on the convex relaxation methods discussed in the previous subsections, we equivalently restate the problem (\ref{eq:problem-original}) as
\begingroup
\allowdisplaybreaks
\begin{subequations} \label{eq:problem-modified}
\begin{align}
    \underset{\boldsymbol{\alpha}, \mathbf{S}, \boldsymbol{\Omega}, \boldsymbol{\tau}, \mathbf{R}} {\mathrm{min.}}\,\,\,
    & \tau_K^{S+W+F,\text{TD}} \!+\! \tau^{W,\text{NO}} \!+\! \max\big\{ \tau^{F, \text{NO}}\!, \tau^{W, \text{CO}} \big\} \!+\! \tau^{F, \text{CO}} \label{eq:problem-original-cost} \\
 \mathrm{s.t. }\,\,\,\,\,\, & \text{Fronthaul time constraints: (\ref{eq:convexified-time-fronthaul-TDMA}), (\ref{eq:convexified-time-fronthaul-NOMA-cooperative})}, \label{eq:problem-modified-time-fronthaul} \\
 & \text{Wireless time constraints: (\ref{eq:convexified-time-wireless-TDMA}), (\ref{eq:convexified-time-wireless-NOMA-cooperative})}, \label{eq:problem-modified-time-wireless} \\
 & \text{Data rate constraints: (\ref{eq:matrix-Lagrange-dual-transform-rate-TDMA}), (\ref{eq:matrix-Lagrange-dual-transform-rate-NOMA-cooperative})}, \label{eq:problem-modified-rate} \\
 & \tau_k^S = \left(\alpha_0 + \alpha_k^{\text{TD}} + \alpha_k^{\text{NO}}\right)\tau^S_{\text{total}}, \, k\in\mathcal{K}, \label{eq:problem-modified-time-sensing-1} \\
 & \tau_k^{S+W,\text{TD}} \geq \max\{\tau_k^S, \tau_{k-1}^{S+W,\text{TD}}\} + \tau_k^{W,\text{TD}}, \, k\in\mathcal{K}, \label{eq:problem-modified-time-sensing-wireless-TDMA} \\
 & \tau_k^{S+W+F, \text{TD}} \geq \label{eq:problem-modified-time-sensing-wireless-fronthaul-TDMA} \\ & \max \big\{ \tau_k^{S+W,\text{TD}}, \tau_{k-1}^{S+W+F, \text{TD}} \big\} + \tau_k^{F, \text{TD}}, \, k\in\mathcal{K}, \nonumber  \\
 & \text{tr}\left(\tilde{\mathbf{S}}_k^{\text{TD}}\tilde{\mathbf{S}}_k^{\text{TD}H}\right) \leq P_U, \, k\in\mathcal{K}, \label{eq:problem-modified-TDMA-power-constraint} \\
 & \text{tr}\left(\tilde{\mathbf{S}}_k^{\text{NO}}\tilde{\mathbf{S}}_k^{\text{NO}H}\right) \leq P_U, \, k\in\mathcal{K}, \label{eq:problem-modified-NOMA-power-constraint} \\
 & \text{tr}\left( \mathbf{E}_k^H\tilde{\mathbf{S}}^{\text{CO}}\tilde{\mathbf{S}}^{\text{CO}H}\mathbf{E}_k \right) \leq P_U, \, k\in\mathcal{K}, \label{eq:problem-modified-cooperative-power-constraint} \\
 & \alpha_0 + \sum\nolimits_{k\in\mathcal{K}} \left(\alpha_k^{\text{TD}} + \alpha_k^{\text{NO}}\right) = 1, \label{eq:problem-modified-alpha-1} \\
 & \alpha_k\geq\alpha_{k-1}, \, k\in\mathcal{K}\setminus\{1\}, \label{eq:problem-modified-alpha-2}
\end{align}
\end{subequations}
\endgroup
with notations $\tilde{\mathbf{S}} = \{\tilde{\mathbf{S}}_k^{\text{TD}}\}_{k\in\mathcal{K}}\cup\{\tilde{\mathbf{S}}_k^{\text{NO}}\}_{k\in\mathcal{K}} \cup \{\tilde{\mathbf{S}}^{\text{CO}}\}$, $\boldsymbol{\theta} = \{\theta_k^{F,\text{TD}}, \theta_k^{W,\text{TD}},\theta_k^{W,\text{NO}}\}_{k\in\mathcal{K}}\cup\{\theta^{F,\text{NO}},\theta^{F,\text{CO}},\theta^{W,\text{CO}}\}$, $\mathbf{G} = \{G_{i,k}^{\text{TD}}\}_{i\in\mathcal{M},k\in\mathcal{K}}\cup\{G_i^{\text{NO}},G_i^{\text{CO}}\}_{i\in\mathcal{M}}$, $\boldsymbol{\Sigma} = \{\boldsymbol{\Sigma}_{i,k}^{\text{TD}}\}_{i\in\mathcal{M},k\in\mathcal{K}}\cup\{\boldsymbol{\Sigma}_i^{\text{NO}}, \boldsymbol{\Sigma}_i^{\text{CO}}\}_{i\in\mathcal{M}}$, $\boldsymbol{\Gamma} = \{\boldsymbol{\Gamma}_k^{\text{TD}},\boldsymbol{\Gamma}_k^{\text{NO}}\}_{k\in\mathcal{K}}\cup\{\boldsymbol{\Gamma}^{\text{CO}}\}$, and $\boldsymbol{\Phi} = \{\boldsymbol{\Phi}_k^{\text{TD}},\boldsymbol{\Phi}_k^{\text{NO}}\}_{k\in\mathcal{K}}\cup\{\boldsymbol{\Phi}^{\text{CO}}\}$.

While inherently non-convex, an appealing characteristic of the problem (\ref{eq:problem-modified}) is that it becomes a convex one when the variables $\{\boldsymbol{\theta}, \boldsymbol{\mathbf{G}}, \boldsymbol{\Sigma}, \boldsymbol{\Gamma}, \boldsymbol{\Phi}\}$ are treated as constants.
Furthermore, closed-form solutions for the optimal values of $\{\boldsymbol{\theta}, \boldsymbol{\mathbf{G}}, \boldsymbol{\Sigma}, \boldsymbol{\Gamma}, \boldsymbol{\Phi}\}$, while keeping the primary variables fixed, are derived as expressed in equations (\ref{eq:optimal-auxiliary-time-fronthaul-TDMA}), (\ref{eq:optimal-auxiliary-time-fronthaul-NOMA-cooperative}), (\ref{eq:optimal-auxiliary-time-wireless-TDMA}), (\ref{eq:optimal-auxiliary-time-wireless-NOMA-cooperative}), (\ref{eq:optimal-auxiliary-rate-TDMA}), and (\ref{eq:optimal-auxiliary-rate-NOMA-cooperative}).
Therefore, we propose an alternating optimization approach which alternately optimizes the primary variables $\{\boldsymbol{\alpha}, \tilde{\mathbf{S}}, \boldsymbol{\Omega}, \boldsymbol{\tau}, \mathbf{R}\}$ and auxiliary variables $\{\boldsymbol{\theta}, \boldsymbol{\mathbf{G}}, \boldsymbol{\Sigma}, \boldsymbol{\Gamma}, \boldsymbol{\Phi}\}$.
With this approach, monotonically decreasing objective values can be obtained, consequently leading to a locally optimal solution.
The detailed algorithm is described in Algorithm 1.
Due to the inherent non-convexity of problem (\ref{eq:problem-modified}), Algorithm 1 cannot guarantee a globally optimal solution. To increase the probability of finding a global solution, the algorithm can be run with multiple randomly chosen initial points, selecting the converged solution with the shortest completion time.
\begin{algorithm}
\caption{Proposed alternating optimization algorithm}

\textbf{\footnotesize{}1}~\textbf{initialize:}

\textbf{\footnotesize{}2}~Set $\{\boldsymbol{\alpha}, \tilde{\mathbf{S}}\}$ to arbitrary values that satisfy the constraints (\ref{eq:problem-modified-TDMA-power-constraint})-(\ref{eq:problem-modified-alpha-2}).

\textbf{\footnotesize{}3}~Set $\boldsymbol{\Omega}$ to arbitrary positive semidefinite matrices.

\textbf{\footnotesize{}4}~Compute $\mathbf{R}$ according to (\ref{eq:data-rate-TDMA}), (\ref{eq:data-rate-NOMA}), and (\ref{eq:data-rate-cooperative-tx}).

\textbf{\footnotesize{}5}~Compute $\boldsymbol{\tau}$ according to (\ref{eq:sensing-time-UAV-k}), (\ref{eq:time-sensing-plus-TDMA-each-UAV}), (\ref{eq:time-sensing-plus-TDMA-plus-FH-each-UAV}), (\ref{eq:total-completion-time}), (\ref{eq:time-TDMA-fronthaul}), (\ref{eq:time-TDMA-wireless}), (\ref{eq:time-NOMA-fronthaul}), (\ref{eq:time-NOMA-wireless}), (\ref{eq:time-cooperative-fronthaul}), and (\ref{eq:time-wireless-cooperative-tx}), and store $\tau_{\text{total}}^{\text{old}} \leftarrow \tau_{\text{total}}$.

\textbf{\footnotesize{}6}~\textbf{repeat}

\textbf{\footnotesize{}7}~Update $\{\boldsymbol{\theta}, \boldsymbol{\mathbf{G}}, \boldsymbol{\Sigma}, \boldsymbol{\Gamma}, \boldsymbol{\Phi}\}$ according to (\ref{eq:optimal-auxiliary-time-fronthaul-TDMA}), (\ref{eq:optimal-auxiliary-time-fronthaul-NOMA-cooperative}), (\ref{eq:optimal-auxiliary-time-wireless-TDMA}), (\ref{eq:optimal-auxiliary-time-wireless-NOMA-cooperative}), (\ref{eq:optimal-auxiliary-rate-TDMA}), and (\ref{eq:optimal-auxiliary-rate-NOMA-cooperative}).

\textbf{\footnotesize{}8}~Update $\{\boldsymbol{\alpha}, \tilde{\mathbf{S}}, \boldsymbol{\Omega}, \boldsymbol{\tau}, \mathbf{R}\}$ as a solution of the convex problem obtained by fixing $\{\boldsymbol{\theta}, \boldsymbol{\mathbf{G}}, \boldsymbol{\Sigma}, \boldsymbol{\Gamma}, \boldsymbol{\Phi}\}$ in (\ref{eq:problem-modified}), and store \\$\tau_{\text{total}}^{\text{new}} \leftarrow \tau_{\text{total}}$.

\textbf{\footnotesize{}9}~\textbf{until} $|\tau_{\text{total}}^{\text{new}} - \tau_{\text{total}}^{\text{old}}|\leq\epsilon$ (Otherwise, set $\tau_{\text{total}}^{\text{old}} \leftarrow \tau_{\text{total}}^{\text{new}}$);
\end{algorithm}

\subsection{Complexity Discussion} \label{sub:complexity}

The complexity of Algorithm 1 is given by the complexity of each iteration, i.e., Steps 7 and 8, multiplied by the number of iterations required for convergence. For the latter, we will provide numerical results showing that the proposed algorithm converges within a few iterations. The former, i.e., the complexity of each iteration, is dominated by the complexity required to solve the convex problem at Step 8, i.e., the convex problem obtained by fixing $\{\boldsymbol{\theta}, \boldsymbol{\mathbf{G}}, \boldsymbol{\Sigma}, \boldsymbol{\Gamma}, \boldsymbol{\Phi}\}$ in problem (\ref{eq:problem-modified}). According to \cite[p. 4]{BTal:19}, the worst-case complexity of solving a generic convex optimization problem is given as $\mathcal{O} \left( n_v\left( n_v^3 + n_c \right) \log(1/\epsilon) \right)$, where $n_v$ indicates the number of optimization variables, $n_c$ denotes the number of arithmetic operations required to compute the objective and constraint function, and $\epsilon$ represents the desired error tolerance level. The numbers $n_v$ and $n_c$ for the convex problem tackled in Step 8 are given as $n_v = \mathcal{O} \left( K\left( Kn_U^2+Mn_A^2\right)\right)$ and $n_c = \mathcal{O} \left( KMn_A \left( n_A^2+Kn_U^2 + Mn_Un_A\right)\right)$, respectively.

\subsection{Alternative Orthogonal Multiple Access Technique} \label{sub:other-multiple-access}

Instead of TDMA, one can consider applying the orthogonal frequency-division multiple access (OFDMA) technique for orthogonal transmission of the first private submessages. In this approach, the private sensory data of each UAV is divided into two submessages, which are sequentially transmitted using OFDMA and NOMA modes.
Although the overall procedure is similar to the proposed TDMA-NOMA transmission, the key difference is that when the UAVs send their first submessages for private sensing results with the OFDMA mode, they can transmit in parallel, not sequentially. This leads to the expression for the time $\tau_k^{S+W,\text{OFD}}$ required for UAV $k$ to complete its sensing task and its OFDMA transmission as
\begin{align}
    \tau_k^{S+W,\text{OFD}} = \tau_k^S + \tau_k^{W,\text{OFD}},
\end{align}
where $\tau_k^{W,\text{OFD}}$ denotes the time spent by UAV $k$ to transmit in the OFDMA mode. This differs from the time expression $\tau_k^{S+W,\text{TD}} = \max{ \{ \tau_k^S, \tau_{k-1}^{S+W,\text{TD}} } \} + \tau_k^{W,\text{TD}}$ for the TDMA mode, which reflects that UAV $k$ can start its TDMA transmission only after both its sensing task and the TDMA transmission of UAV $k-1$ are completed.

While OFDMA allows for parallel transmission among UAVs unlike TDMA, the wireless transmission time of each UAV is shorter for TDMA transmission than for OFDMA, i.e., $\tau_k^{W,\text{TD}} < \tau_k^{W,\text{OFD}}$. This is because, with OFDMA, each UAV uses only a portion of the available uplink bandwidth, while TDMA allows each UAV to occupy all the available bandwidth thanks to the separation in the time domain.
Therefore, for a fair comparison between the TDMA and OFDMA techniques for orthogonal private data transmission, we should address the same problem of minimizing the total mission completion time under the hybrid OFDMA-NOMA-cooperative transmission strategy. This can be achieved by applying the proposed algorithm with appropriate modifications.
A thorough comparison among different multiple access techniques with comprehensive analysis is left for future work.

\section{Numerical Results} \label{sec:numerical-results}

\begin{figure}
\centering\includegraphics[width=0.7\linewidth]{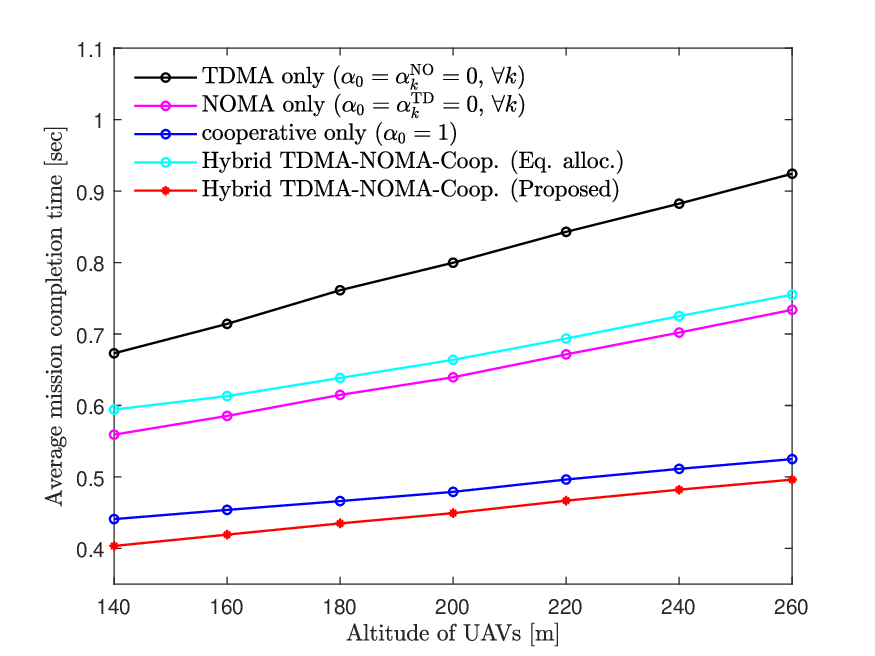}
\vspace{-3mm}
\caption{\small Average mission completion time versus the altitude of UAVs ($K=4$, $M=2$, $n_U=n_A=2$, $P_U/\sigma_z^2=15$ dB, $C_F=1$  Gbps, $b_\text{total} = 200$ Mbits, $\tau_{\text{total}}^S = 0.1$ sec).} \label{fig:MCT-Altitude}
\end{figure}

In this section, we present numerical results to validate the performance enhancements achieved by the proposed hybrid TDMA-NOMA-cooperative transmission strategy for multi-UAV collaborative sensing in cell-free MIMO networks, as compared to conventional baseline transmission schemes.
We assume that the UAVs are located randomly within a circular area of radius 50 m with the purpose of collaborative sensing of the region of interest at a fixed altitude of 200 m, except in Fig. \ref{fig:MCT-Altitude} where the impact of varying altitude is examined. To facilitate connectivity between the UAVs and the CS, the APs are positioned randomly within a circular area surrounding the region of interest, with a radius of 200 m and ground-level height.
For the air-to-ground link from the UAVs to APs, we adopt the Rician fading channel model \cite{Xu:TC23} under the assumption that both the UAVs and the APs are equipped with uniform linear arrays (ULAs).
The ULAs of the APs are assumed to be aligned in parallel, while the orientations of the UAVs' ULAs are random.
The channel matrix  $\mathbf{H}_{i,k}\in\mathbb{C}^{n_A\times{n_U}}$ between the UAV $k$ and AP $i$ is then given as
\begin{align}
    \mathbf{H}_{i,k} = \sqrt{\beta_{i,k}} \left( \sqrt{\frac{\kappa}{\kappa+1}}\mathbf{H}_{i,k}^\text{LoS} + \sqrt{\frac{1}{\kappa+1}}\mathbf{H}_{i,k}^\text{NLoS}\right),
    \label{eq:channel}
\end{align}
where $\beta_{i,k}=\beta_0(d_{i,k}/d_0)^{-3}$ is the pathloss between AP $i$ and UAV $k$ with $d_{i,k}$ being the distance between the AP $i$ and UAV $k$. We set the reference distance and pathloss as $d_0=30$ m and $\beta_0=10$. $\kappa$ represents the Rician K-factor set to $\kappa = 100$, and $\mathbf{H}_{i,k}^\text{LoS}\in\mathbb{C}^{n_A\times{n_U}}$ and $\mathbf{H}_{i,k}^\text{NLoS}\in\mathbb{C}^{n_A\times{n_U}}$ are the LoS and non-LoS (NLoS) components, respectively. The elements of $\mathbf{H}_{i,k}^\text{NLoS}$ are independent and identically distributed (i.i.d.) as $\mathcal{CN}(0,1)$, and the LoS component $\mathbf{H}_{i,k}^\text{LoS}$ is modelled as \cite{Xu:TC23}
\begin{align}
    \mathbf{H}_{i,k}^\text{LoS} = \sqrt{n_U n_A}\, \tilde{\mathbf{h}}_\text{R}^\text{LoS} \left( \theta_{i,k}^\text{AoA}\right )\left(\tilde{\mathbf{h}}_\text{T}^\text{LoS} \left( \theta_{i,k}^\text{AoD}\right )\right)^H ,
    \label{eq:channel-LoS-component}
\end{align}
where $\tilde{\mathbf{h}}_\text{R}^\text{LoS} \left( \theta_{i,k}^\text{AoA}\right )\in\mathbb{C}^{n_A\times 1}$ and $\tilde{\mathbf{h}}_\text{T}^\text{LoS} \left( \theta_{i,k}^\text{AoD}\right)\in\mathbb{C}^{n_U\times 1}$ represent the receive and transmit array responses at the AP $i$ and UAV $k$, respectively, defined as
\begin{subequations}\label{array response}
\begin{align}
    &\tilde{\mathbf{h}}_\text{R}^\text{LoS} \left( \theta_{i,k}^\text{AoA}\right ) = \frac{1}{\sqrt{n_A}} \left[ 1\,\,e^{-j\frac{2 \pi d_c}{\lambda_c}\sin{\left(\theta_{i,k}^\text{AoA}\right)}}\, \cdots \,e^{-j\frac{2 \pi \left( n_A-1\right)d_c}{\lambda_c}\sin{\left(\theta_{i,k}^\text{AoA}\right)}}\right]^T, \label{eq:receive array response} \\
    &\tilde{\mathbf{h}}_\text{T}^\text{LoS} \left( \theta_{i,k}^\text{AoD}\right ) = \frac{1}{\sqrt{n_U}} \left[ 1\,\, e^{-j\frac{2 \pi d_c}{\lambda_c}\sin{\left(\theta_{i,k}^\text{AoD}\right)}}\, \cdots\,e^{-j\frac{2 \pi \left( n_U-1\right)d_c}{\lambda_c}\sin{\left(\theta_{i,k}^\text{AoD}\right)}}\right]^T. \label{eq:transmit array response}
\end{align}
\end{subequations}
Here $\theta_{i,k}^\text{AoA}$ and $\theta_{i,k}^\text{AoA}$ represent the angle-of-arrival (AoA) and the angle-of-departure (AoD), respectively, $\lambda_c$ is the carrier wavelength, and $d_c$ represents the distance between adjacent antenna elements set to $d_c = f_c / (2c)$.
Throughout the simulation, the bandwidth and the carrier frequency  are set to be $B=100$ MHz and $f_c=2.5$ GHz, respectively.
When optimizing $\{\boldsymbol{\alpha}, \tilde{\mathbf{S}}, \boldsymbol{\Omega}, \boldsymbol{\tau}, \mathbf{R}\}$ for fixed $\{\boldsymbol{\theta}, \mathbf{G}, \boldsymbol{\Sigma}, \boldsymbol{\Gamma}, \boldsymbol{\Phi}\}$ in  Step 8 of Algorithm 1, we adopt the CVX software tool \cite{Grant:CVX20}.
For the SIC decoding in the NOMA phase, we randomly set the decoding order $\pi$.

\begin{figure}
\centering\includegraphics[width=0.7\linewidth]{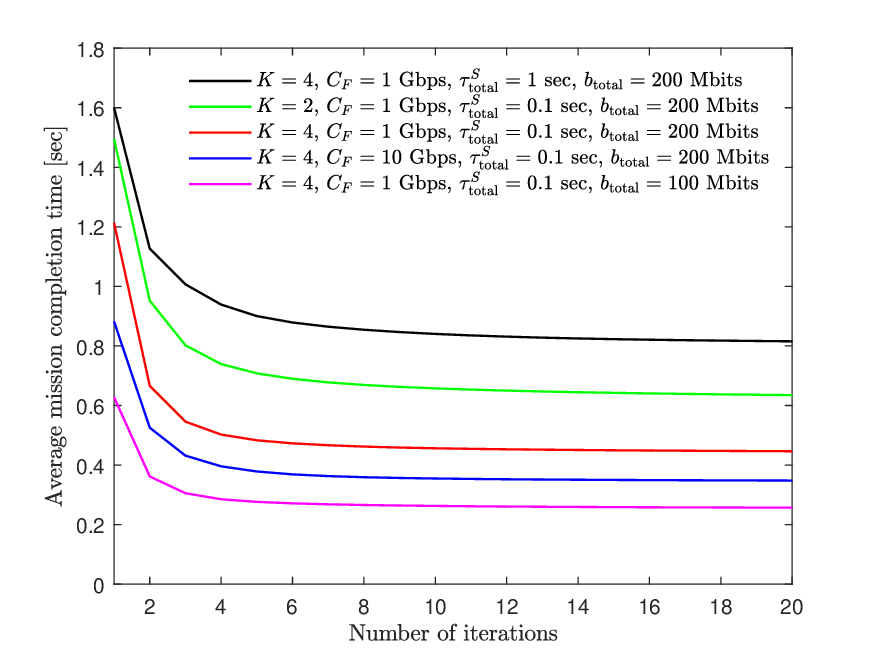}
\vspace{-4mm}
\caption{\small Average completion time versus the number of iterations ($M=2$, $n_U=n_A=2$, $P_U/\sigma_z^2=15$ dB, $K\in\{2,4\}$, $C_F\in\{1,10\}$ Gbps, $\tau_{\text{total}}^S\in\{0.1, 1\}$ sec, and $b_{\text{total}}\in\{100,200\}$ Mbits).} \label{fig:graph-vs-iterations}
\end{figure}

In Fig. \ref{fig:graph-vs-iterations}, we numerically check the convergence speed of the proposed algorithm by plotting the average mission completion time $\tau_{\text{total}}$ versus the number of iterations for a collaborative sensing system with $M=2$, $n_U=n_A=2$, $P_U/\sigma_z^2=15$ dB, $K\in\{2,4\}$, $C_F\in\{1,10\}$ Gbps, $\tau_{\text{total}}^S\in\{0.1, 1\}$ sec, and $b_{\text{total}}\in\{100,200\}$ Mbits.
The graph demonstrates that the proposed algorithm converges within only a few iterations for all simulated scenarios, validating its rapid convergence.

\begin{figure}
\centering\includegraphics[width=0.7\linewidth]{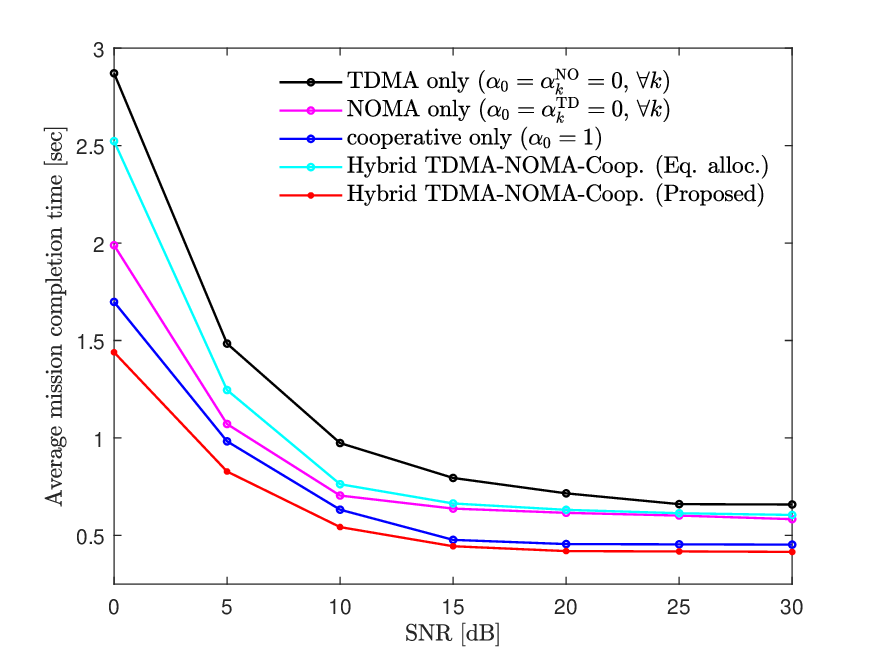}
\vspace{-4mm}
\caption{\small Average mission completion time versus the SNR $P_U/\sigma_z^2$ ($K=4$, $M=2$, $n_U=n_A=2$, $C_F = 1$ Gbps, $\tau_{\text{total}}^S = 0.1$ sec, $b_{\text{total}}= 200$ Mbits).} \label{fig:MCT-SNR}
\end{figure}

In Fig. \ref{fig:MCT-SNR}, the average mission completion time is depicted while increasing the SNR level $P_U/ \sigma_z^2$ for a collaborative sensing system with $K=4$, $M=2$, $n_U=n_A=2$, $C_F = 1$ Gbps, $\tau_{\text{total}}^S = 0.1$ sec, and $b_{\text{total}} = 200$ Mbits.
We evaluate the performance of the proposed optimized hybrid TDMA-NOMA-cooperative transmission against several baseline schemes:
\textbf{\textit{i) TDMA only}} ($\alpha_0=0$ and $\alpha_k^{\text{NO}}=0$, $\forall k \in\mathcal{K}$);
\textbf{\textit{ii) NOMA only}} ($\alpha_0=0$ and $\alpha_k^{\text{TD}}=0$, $\forall k\in\mathcal{K}$);
\textbf{\textit{iii) Cooperative only}} ($\alpha_0=1$);
\textbf{\textit{iv) Hybrid TDMA-NOMA-cooperative strategy with equal task allocation}} ($\alpha_0=\alpha_k^{\text{TD}}=\alpha_k^{\text{NO}}=1/(2K+1)$, $\forall k\in\mathcal{K}$).
The figure reveals that the hybrid transmission strategy yields a worse completion time compared to NOMA and cooperative transmission schemes when task allocation variables $\boldsymbol{\alpha}$ are uniformly distributed (i.e., equal task allocation). However, when the hybrid transmission strategy operates with optimized allocation variables $\boldsymbol{\alpha}$ according to Algorithm 1, it outperforms all baseline schemes in terms of completion time.
Notably, as the SNR becomes sufficiently large, the cooperative transmission scheme dominates all other baseline schemes, causing the proposed optimized hybrid scheme to converge towards the pure cooperative transmission scheme.

\begin{figure*}
\centering

\captionsetup[subfigure]{labelformat=empty}

\subfloat[Mission completion time ($P_U/\sigma_z^2=0$ dB)]{\includegraphics[width=7.0cm] {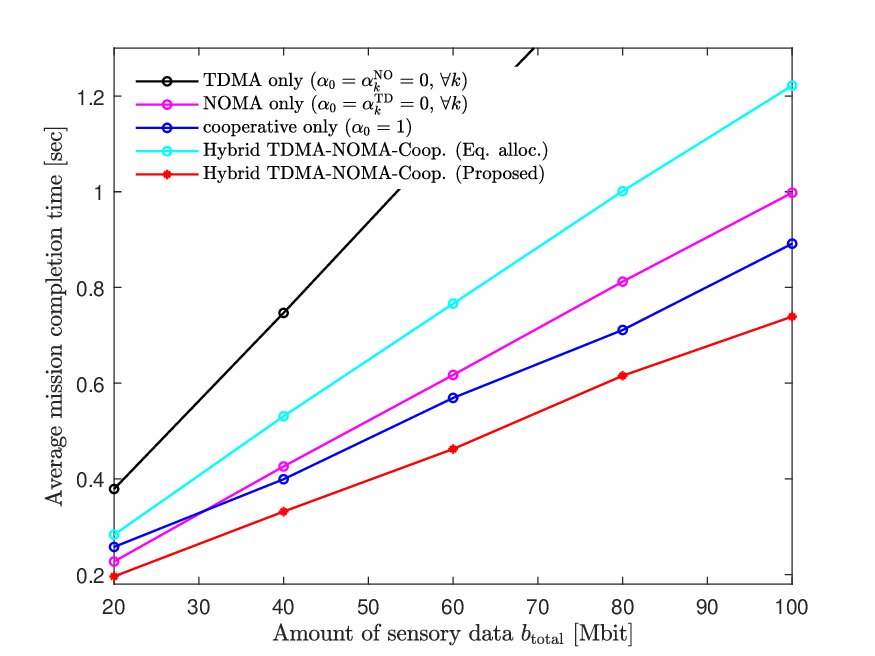}}\hfil
\subfloat[Task allocation ($P_U/\sigma_z^2=0$ dB)]{\includegraphics[width=7.0cm]{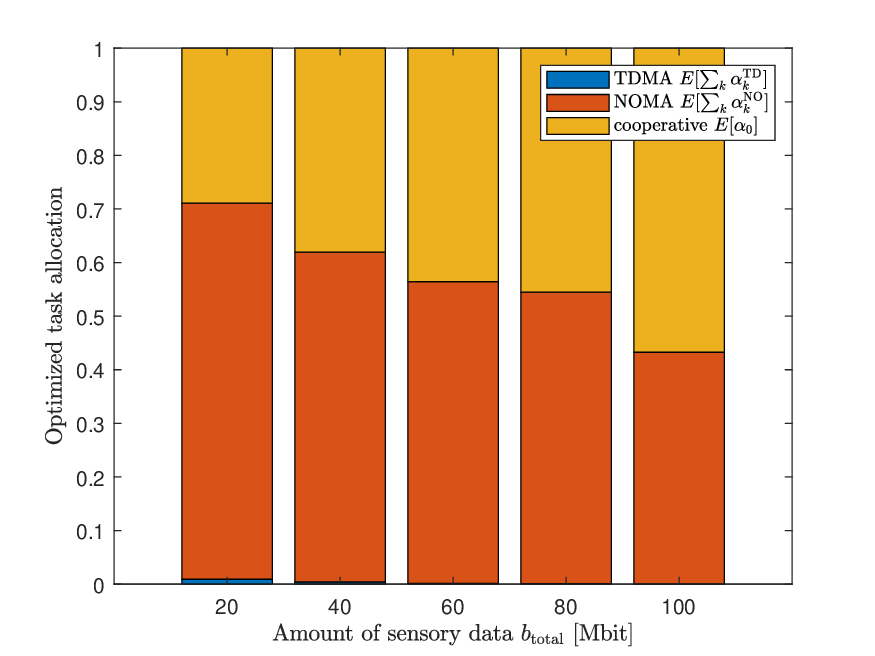}}

\subfloat[Mission completion time ($P_U/\sigma_z^2=30$ dB)]{\includegraphics[width=7.0cm]{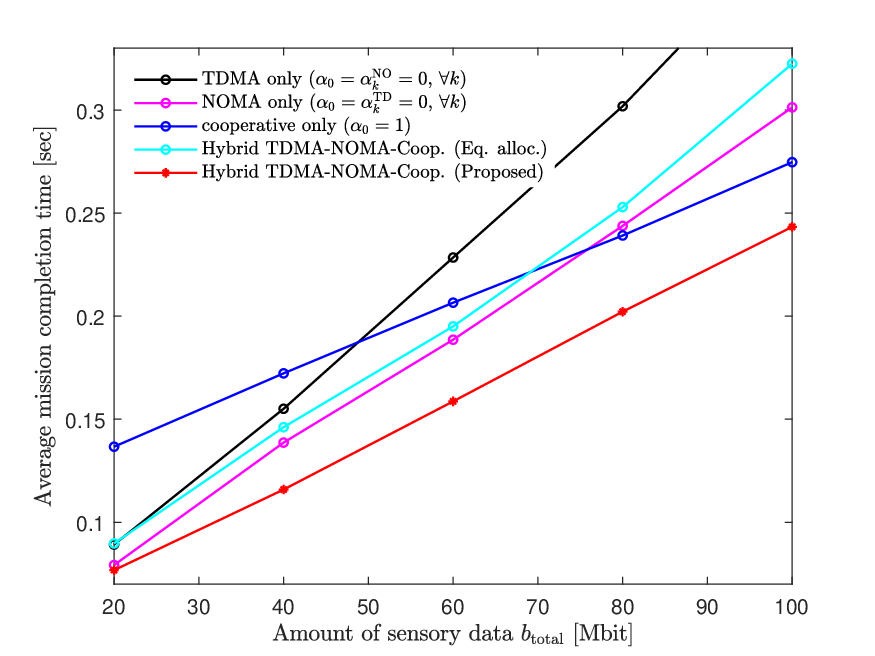}}\hfil
\subfloat[Task allocation ($P_U/\sigma_z^2=30$ dB)]{\includegraphics[width=7.0cm]{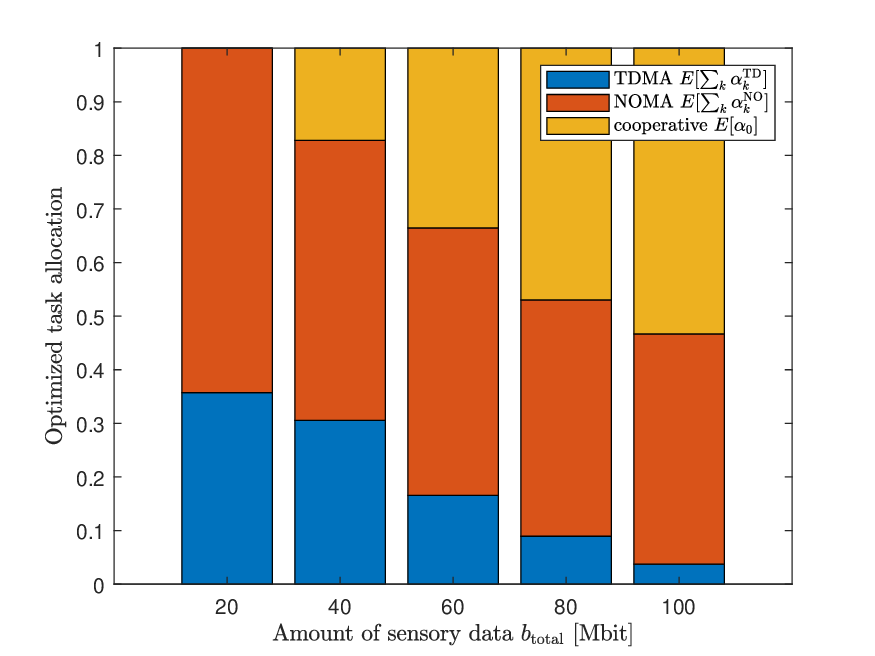}}
\caption{\small Average mission completion time versus the amount of sensory data $b_{\text{total}}$ ($K=4$, $M=2$, $n_U=n_A=2$, $C_F = 1$ Gbps, $P_U/\sigma_z^2\in\{0, 30\}$ dB, $\tau_{\text{total}}^S = 0.1$ sec).}\label{fig:MCT-bits}
\vspace{-4mm}
\end{figure*}

In Fig. \ref{fig:MCT-bits}, we delve into the effects of the total amount of sensory data $b_{\text{total}}$ on the mission completion time by presenting the average completion time and the optimized task allocation variables relative to $b_{\text{total}}$, considering a system with $K=4$, $M=2$, $n_U=n_A=2$, $C_F = 1$ Gbps, $P_U/\sigma_z^2\in\{0, 30\}$ dB, and $\tau_{\text{total}}^S = 0.1$ sec.
The cooperative transmission strategy leverages the array gain across UAVs, thereby reducing transmission time over the wireless link compared to the TDMA and NOMA schemes. Instead, it does introduce an additional delay in the execution of sensing tasks, as UAVs should undertake overlapping tasks.
In Fig. \ref{fig:MCT-bits}, we observe that as $b_{\text{total}}$ increases, the cooperative transmission scheme outperforms the TDMA and NOMA schemes, with the proposed optimized hybrid scheme allocating a larger fraction of tasks to cooperative transmission mode.
This is because the portion of transmission time becomes more prominent in comparison to sensing time when $b_{\text{total}}$  is large.
In high-SNR cases, where TDMA and NOMA schemes exhibit comparable latency to the cooperative transmission scheme, the fraction of the latter is decreased in comparison to low-SNR scenarios.

\begin{figure}
\centering\includegraphics[width=0.7\linewidth]{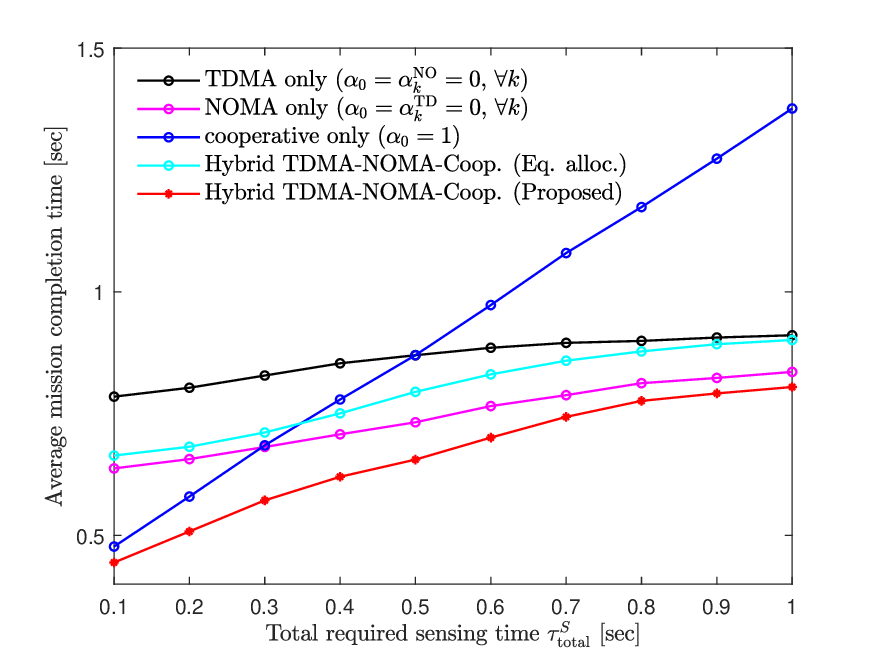}
\vspace{-4mm}
\caption{\small Average mission completion time versus the total required sensing time $\tau_{\text{total}}^S$ ($K=4$, $M=2$, $n_U=n_A=2$, $P_U/\sigma_z^2=15$ dB, $C_F=1$  Gbps, $b_\text{total} = 200$ Mbits).} \label{fig:MCT-Sensingtime}
\end{figure}

In Fig. \ref{fig:MCT-Sensingtime}, the average mission completion time is plotted against the total required sensing time $\tau_{\text{total}}^S$ for a collaborative sensing system configured with $K=4$, $M=2$, $n_U=n_A=2$, $P_U/\sigma_z^2=15$ dB, $C_F=1$ Gbps, and $b_\text{total} = 200$ Mbits.
early demonstrates that as the required sensing time increases, the completion time of the cooperative scheme escalates more rapidly than the other schemes. This phenomenon arises because UAVs undertake the entire sensing mission individually, without dividing the sensing tasks among themselves.
Consequently, crosspoints occur between the cooperative scheme and the other baseline strategies. Notably, the proposed hybrid scheme, that operates with optimized task splitting, achieves significantly lower completion times compared to baseline schemes.

\begin{figure}
\centering\includegraphics[width=0.7\linewidth]{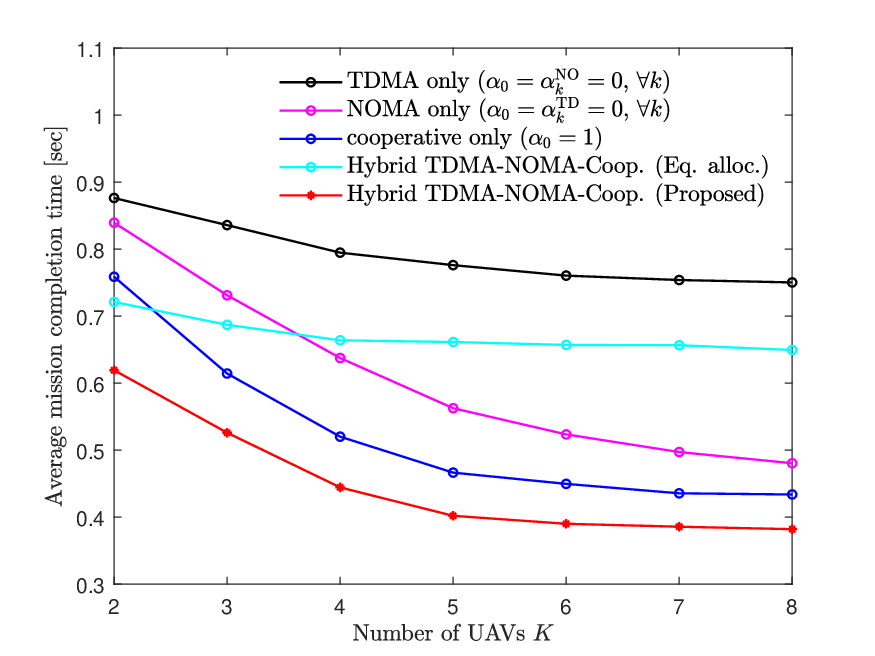}
\vspace{-4mm}
\caption{\small Average mission completion time versus the number of UAVs $K$ ($M=2$, $n_U=n_A=2$, $P_U/\sigma_z^2=15$ dB, $C_F=1$  Gbps, $b_\text{total} = 200$ Mbits, $\tau_{\text{total}}^S = 0.1$ sec).} \label{fig:MCT-UAVs}
\end{figure}

Fig. \ref{fig:MCT-UAVs} illustrates the average mission completion time with respect to the number of UAVs $K$ for a collaborative sensing system with $M=2$, $n_U=n_A=2$, $P_U/\sigma_z^2=15$ dB, $C_F=1$  Gbps, $b_\text{total} = 200$ Mbits, $\tau_{\text{total}}^S = 0.1$ sec.
From the graph, it is evident that increasing the number of UAVs benefits all schemes due to enhanced sensing and communication capabilities.
However, the performance limitation of the TDMA scheme compared to other schemes becomes more pronounced with an increased $K$, as it does not leverage the spatial multiplexing gain achievable in the cell-free MIMO network with cooperating APs.
Additionally, it is worth highlighting that the proposed hybrid scheme demonstrates notable performance improvements regardless of the number of UAVs.

\begin{figure}
\centering\includegraphics[width=0.7\linewidth]{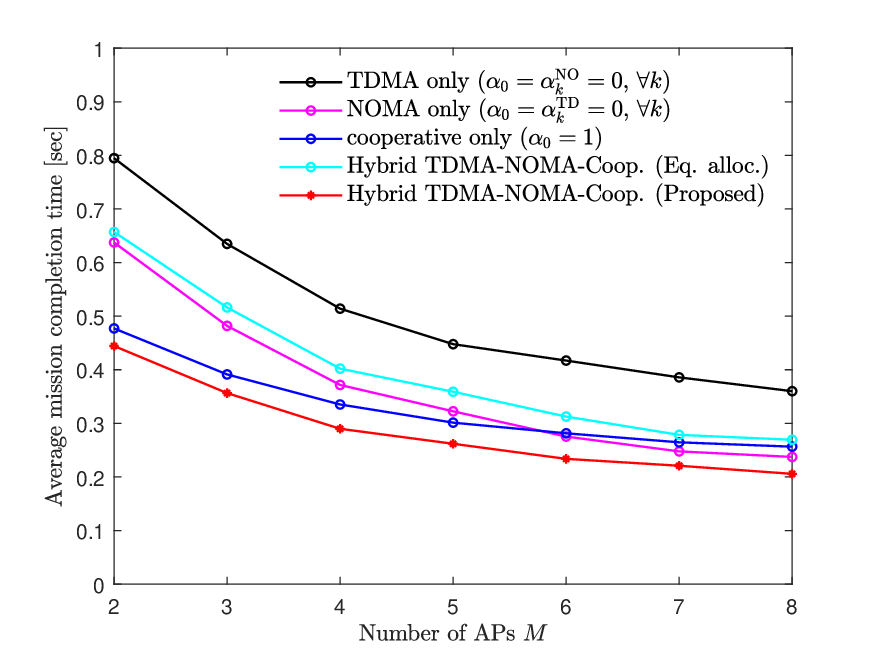}
\vspace{-3mm}
\caption{\small Average mission completion time versus the number of APs $M$ ($K=2$, $n_U=n_A=2$, $P_U/\sigma_z^2=15$ dB, $C_F=1$  Gbps, $b_\text{total} = 200$ Mbits, $\tau_{\text{total}}^S = 0.1$ sec).} \label{fig:MCT-APs}
\end{figure}

In Fig. \ref{fig:MCT-APs}, we depict the average mission completion time against the number of APs $M$ for a collaborative sensing system with $K=2$, $n_U=n_A=2$, $P_U/\sigma_z^2=15$ dB, $C_F=1$  Gbps, $b_\text{total} = 200$ Mbits, $\tau_{\text{total}}^S = 0.1$ sec.
Comparing the performance of the NOMA and cooperative schemes, we observe that with a larger number of APs $M$, the inter-UAV interference signals of the NOMA scheme can be better mitigated, leading to superior performance compared to the cooperative scheme when a sufficient number of APs are present.
Furthermore, the performance improvement of the proposed hybrid scheme is most noticeable with an intermediate number of APs, and its performance converges towards that of the cooperative or NOMA scheme when $M$ is small or large, respectively.

\begin{figure}
\centering\includegraphics[width=0.7\linewidth]{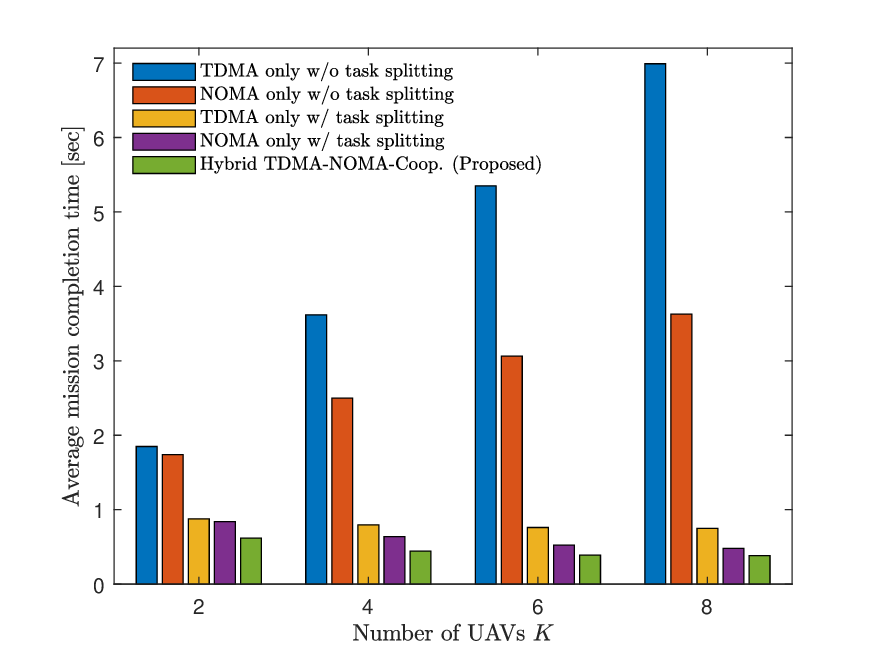}
\vspace{-4mm}
\caption{\small Average mission completion time versus the number of UAVs $K$ ($M=2$, $n_U=n_A=2$, $P_U/\sigma_z^2=15$ dB, $C_F=1$  Gbps, $b_\text{total} = 200$ Mbits, $\tau_{\text{total}}^S = 0.1$ sec).} \label{fig:MCT-tasksplitting}
\end{figure}

In Fig. \ref{fig:MCT-tasksplitting}, we highlight the significance of task splitting when applying either TDMA or NOMA transmission strategies for sensing data transmission.
The figure presents the average mission completion time as a function of the number of UAVs $K$ in a collaborative sensing system with $M=2$, $n_U=n_A=2$, $P_U/\sigma_z^2=15$ dB, $C_F=1$  Gbps, $b_\text{total} = 200$ Mbits, and $\tau_{\text{total}}^S = 0.1$ sec.
We compare the performance of TDMA and NOMA schemes, both with and without task splitting. The figure shows that the optimized task splitting accelerates the sensing and transmission processes for both TDMA and NOMA, with performance gains increasing as the number of UAVs $K$ grows.
Additionally, we compare these with the proposed hybrid TDMA-NOMA-cooperative scheme, which achieves the shortest mission completion time across all simulated configurations.

Fig. \ref{fig:MCT-Altitude} illustrates the average mission completion time against the different heights of UAVs for a collaborative sensing system with $K=4$, $M=2$, $n_U=n_A=2$, $P_U/\sigma_z^2=15$ dB, $C_F=1$  Gbps, $b_\text{total} = 200$ Mbits, and $\tau_{\text{total}}^S = 0.1$ sec.
As the UAVs' altitude increases, the mission completion time grows for all schemes due to the increased pathloss between the UAVs and APs. Notably, the cooperative scheme and the proposed hybrid TDMA-NOMA-cooperative scheme exhibit the smallest increase in mission completion time with respect to altitude. This aligns with the observation in Fig. \ref{fig:MCT-SNR}, where these two schemes also demonstrate the slowest growth in mission completion time as the SNR level decreases.

\section{Conclusion} \label{sec:conclusion}

We have investigated a multi-UAV sensing system that integrates both common and private sensing tasks and reports the sensory data within a cell-free MIMO network. To efficiently utilize the wireless channel from UAVs to APs, we have employed a hybrid transmission strategy, encompassing TDMA, NOMA, and cooperative transmission.
We addressed the problem of jointly optimizing task splitting ratios and hybrid TDMA-NOMA-cooperative transmission strategies to minimize mission completion time.
To tackle the non-convex nature of the problem, we have reformulated the problem using the FP and matrix FP techniques and derived an alternating optimization algorithm that achieves monotonically decreasing completion time values.
Through extensive numerical results, we have demonstrated the efficacy of the proposed approach in expediting sensing mission task completion.
As a future research direction, we envision optimizing the trajectories of UAVs \cite{Jeong:TVT18} within the framework of the proposed multi-UAV sensing system. Additionally, we aim to develop a joint signal and CSI compression approach to enhance the robustness of the proposed multi-UAV system against CSI imperfections \cite{Kang:TWC14}.
Furthermore, a comprehensive comparison of different multiple access techniques such as TDMA, OFDMA, NOMA, and RSMA for the wireless transmission of sensory data would be highly valuable.

\end{document}